\documentclass[%
reprint,
superscriptaddress,
 showpacs,preprintnumbers,
 bibnotes,
 amsmath,amssymb,
 aps,
 prb,
floatfix
]{revtex4-1}

\usepackage{graphicx}
\usepackage{dcolumn}
\usepackage{bm}
\usepackage{epstopdf}
\usepackage{color}

\usepackage{color}
\usepackage{float}
\usepackage{amsfonts}
\usepackage{mathtools}

\begin{document}


\title{Valley qubit in gated MoS$_2$ monolayer quantum dot}
\author{J. Paw\l{}owski}
\email[]{jaroslaw.pawlowski@pwr.edu.pl}
\affiliation{
Department of Theoretical  Physics, Faculty of Fundamental Problems of Technology, Wroc\l{}aw University of Science and Technology, Wybrze\.{z}e Wyspia\'{n}skiego 27, 50-370 Wroc\l{}aw, Poland}
\author{D. \.Zebrowski}
\affiliation{
	Faculty of Physics and Applied Computer Science,
	AGH University of Science and Technology, Krak\'{o}w, Poland}
\author{S. Bednarek}
\affiliation{
	Faculty of Physics and Applied Computer Science,
	AGH University of Science and Technology, Krak\'{o}w, Poland}

\date{\today}

\begin{abstract}
The aim of presented research is to design a nanodevice, based on a MoS$_2$ monolayer, performing operations on a well-defined valley qubit. We show how to confine an electron in a gate induced quantum dot within the monolayer, and to perform the NOT operation on its valley degree of freedom. The operations are carried out all electrically via modulation of the confinement potential by oscillating voltages applied to the local gates. Such quantum dot structure is modeled realistically. Through these simulations we investigate the possibility of realization of a valley qubit in analogy with a realization of the spin qubit. 
We accurately model the potential inside the nanodevice accounting for proper boundary conditions on the gates and space-dependent materials permittivity by solving the generalized Poisson's equation. The time-evolution of the system is supported by realistic self-consistent Poisson-Schr\"odinger tight-binding calculations. The tight-binding calculations are further confirmed by simulations within the effective continuum model.
\vspace{1.2cm}
\end{abstract}


\maketitle


\section{Introduction}
Structures consisting of several atomic layers are modern materials which may find applications in devices for quantum computing or quantum cryptography. Two-dimensional layers exhibit quite different properties than bulk crystals. A truly groundbreaking experiment was isolation and examining of graphene\cite{nobel,nobell,nobel1}.

Graphene is characterized by a very high carrier mobility at room temperature\cite{mob}, which makes it a perfect material for applications in electronics. High carrier mobility translates into ultra-fast transistors for digital electronics. However, the absence of a band gap makes graphene field effect transistors hard to control/turn-off by an electric field. Also the linear dispersion relation makes the electrostatic confinement of individual electrons in the quantum dot (QD) structure impossible due to the Klein tunneling\cite{22,27}. Moreover, graphene is of limited usefulness for spintronics because of a rather low spin-orbit coupling\cite{weak}. Currently, modifications of the graphene structure to lift these limitations are being actively researched\cite{s1,s2,s3,s4,s5,s6,s7,s8,s9}.

Another approach is to search for atomic-thin layers of other materials having similar properties to graphene yet without mentioned disadvantages. Ones of such promising materials are monolayers of molybdenum disulfide, tungsten diselenide and others transition metal dichalcogenide monolayers (TMDC)\cite{1,tmdc2,14}.

TMDC are atomic-thin two-dimensional materials and, as opposed to graphene, have direct band gaps between $1.5$-$2.0$~eV in the optical range\cite{1}. Deposition of gates (gating) in nanostructures made of monolayers is experimentally challenging, yet a considerable progress has been made in this field opening a path towards construction of electrostatic quantum dots\cite{24,26,26a,stm,guang1,guang2}. A finite band gap allows for confining carriers in a monolayer in the area between gates by applying voltages to them (with respect to the substrate), and thus creating confinement. The electrons or holes confined in the QD have a spin degree of freedom which can be used to define a quantum bit (qubit). Strong spin-orbit interaction (SOI) in these materials \cite{4} ($0.03$~\AA$^2$ for MoS$_2$, up to $0.2$~\AA$^2$ for WSe$_2$)\cite{4} allows for fast operations on the spin qubit.

Electrons and holes, in these materials, have additional, beside the spin, discrete ,,valley'' degree of freedom\cite{val1,val2} that can be used to define a unit of quantum information, a qubit. The valley degree of freedom originates from the lattice structure. The vectors connecting two neighboring molybdenum atoms form two non-equivalent families: $R_1,R_3,R_5$ and $R_2,R_4,R_6$ with the nearest sulphur neighbour on the left (right) side, as shown in Fig. \ref{fig:2}. This reflects on the reciprocal lattice where in the corners of the first (hexagonal) Brillouin zone, the K points form two non-equivalent families: \textit{K} and \textit{K'}. Note that both the bottom of the conduction band and the top of the valence band are located at the points \textit{K} and \textit{K'} (we have a direct band gap) not at the $\Gamma$ point. These bands form two non-equivalent valleys \textit{K} and \textit{K'} which can be occupied by qubit carriers.

This additional degree of freedom can be used to encode a quantum of information creating a new field of electronics called valleytronics\cite{val1,opt1,opt2,opt3}, in analogous to spintronics, in which interesting ideas for nanodevices using mentioned materials have also been proposed\cite{loss,y}. The aim of presented research is to design a nanodevice based on a MoS$_2$ monolayer able to perform operations on a well-defined valley qubit. 

\section{Nanodevice Model}
\begin{figure}[t]
	\includegraphics[width=8cm]{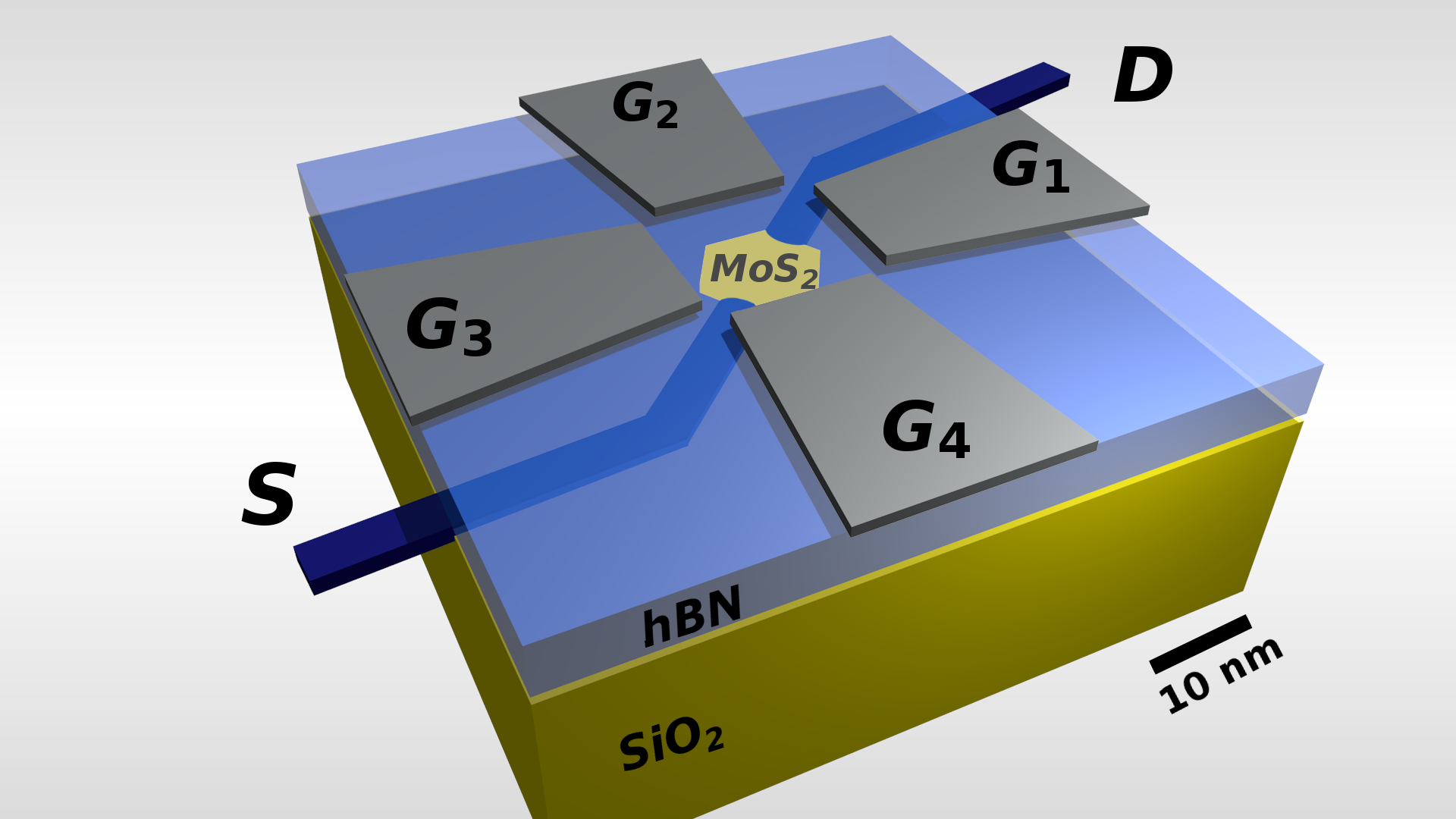}
	\caption{\label{fig:1} The structure of the proposed nanodevice containing a $\mathrm{MoS}_2$ monolayer flake with a layout of local gates responsible for creating the confinement potential of the QD.}
\end{figure}

The proposed nanodevice consists of a MoS$_2$  monolayer flake with nearby gates\cite{ukladanie,26,26a,guang1,guang2}. By applying appropriate voltages to the gates we create confinement in a flake area, which traps a single electron. This way an electrostatic QD is formed within the monolayer. The nanodevice structure is shown in Fig.~\ref{fig:1}.
The lowest layer of the nanodevice is a substrate made of highly donor-doped $\mathrm{Si}^\mathrm{n++}$, which constitutes a bottom gate. Upon this layer we place $20$~nm of an $\mathrm{SiO}_2$ insulator. Next we place a $\mathrm{MoS}_2$ monolayer flake of diameter of at least $10$~nm, necessary to accommodate the confinement of the QD. The entire structure is again covered with an insulator, however this time it is a $5$~nm wide layer of hexagonal boron nitride $\mathrm{hBN}_2$. On top of that sandwiched structure we deposit four gates, symmetrically surrounding the flake. Voltages applied to these gates (relative to the substrate) are used to create the confinement in the flake. To make the potential landscape sharp enough, the distance between the top gates and the substrate should not be excessively long compared to the clearance between the gates (where the flake is placed). The distance defined by the thickness of the insulating layers is $25$~nm as compared to a $15$~nm of clearance. The potential in the entire nanodevice, controlled by the gate voltages, is calculated by solving the generalized Poisson's equation, while the electron states in the flake are described with the tight-binding formalism. 

\subsection{Monolayer}
\begin{figure}[b]
	\includegraphics[width=4.6cm]{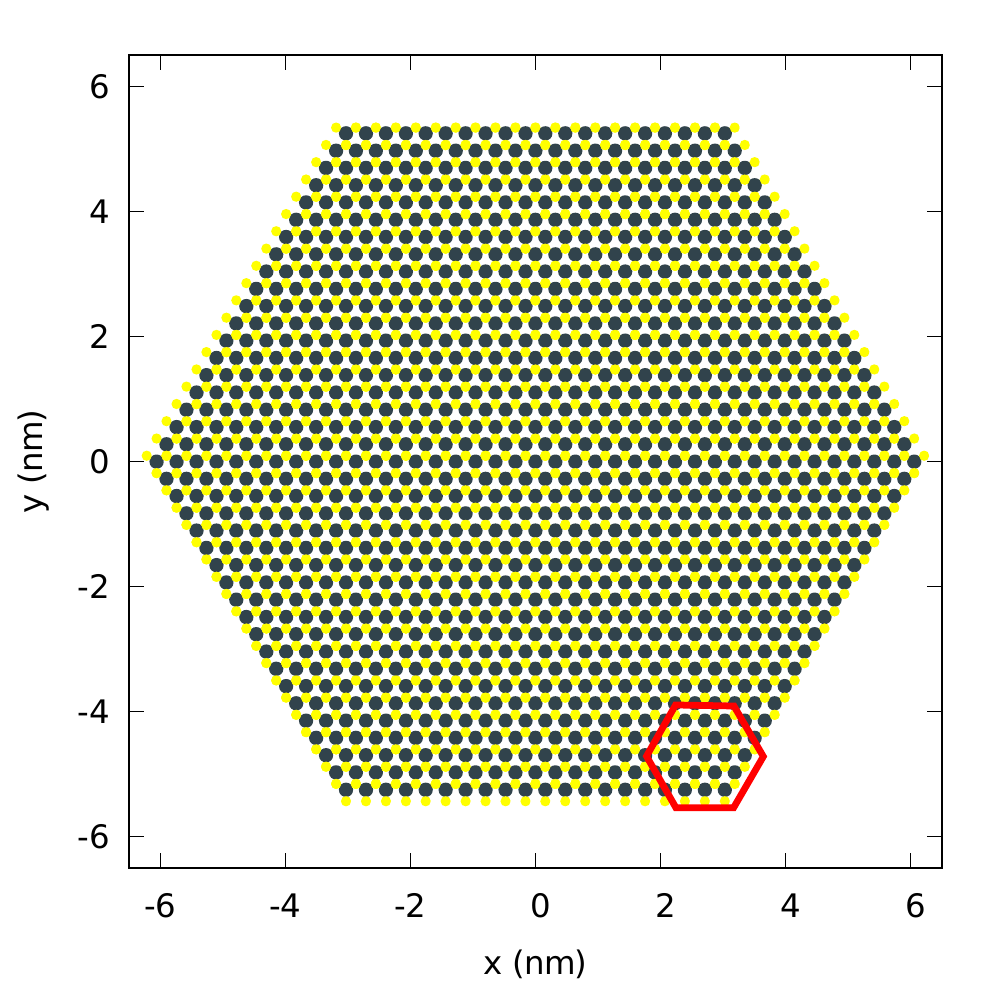}
	\includegraphics[width=3.9cm]{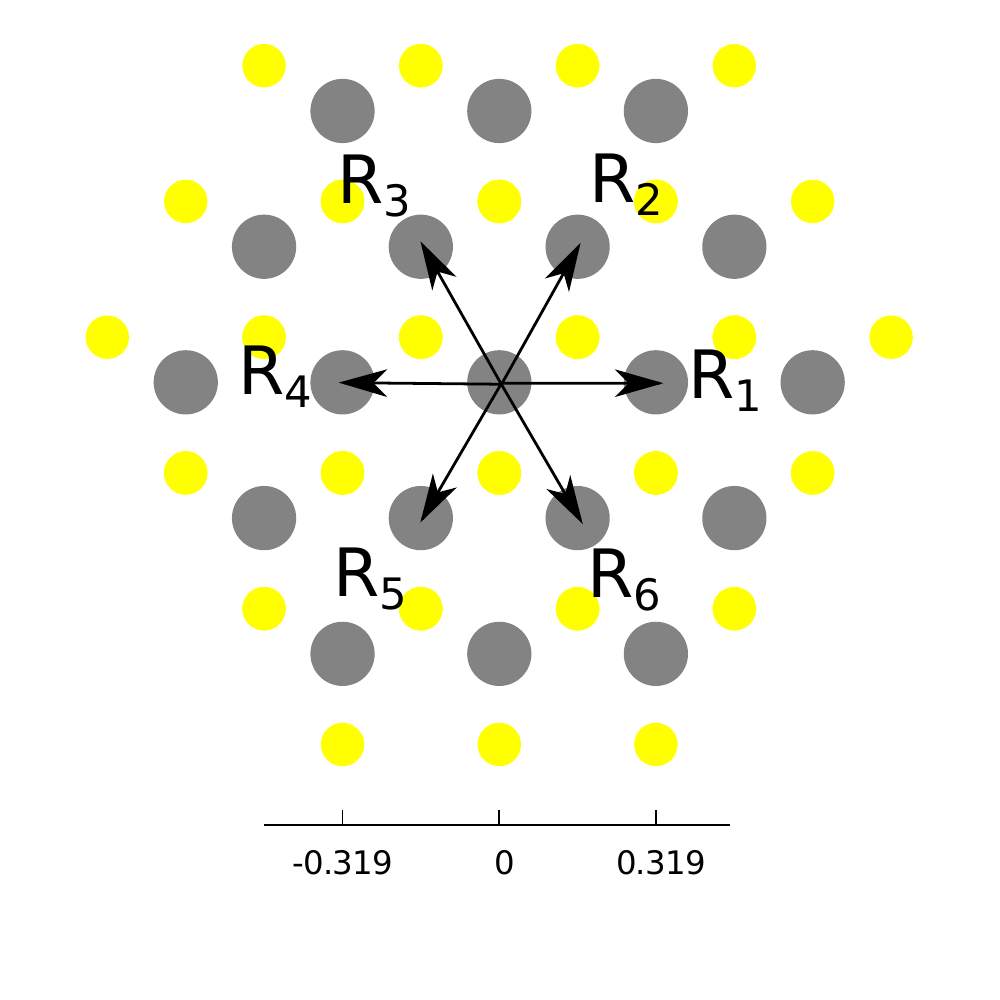}
	\caption{\label{fig:2}  Molybdenum disulfide hexagonal flake with $N_B=20$ Mo atoms on the flake edge (left). The MoS$_2$ monolayer crystal lattice structure: gray Mo and yellow S atoms ($a=0.319$~nm), with marked $R_k$ hopping directions (right).}
\end{figure}

The monolayer flake is made of molybdenum disulfide. It has a hexagonal shape with sides made of $N_B=20$ molybdenum atoms, which results in the side size of $b\simeq20a$, for a crystal lattice constant $a=0.319$~nm. We chose the smallest sufficient size for the flake to speed up the calculations, however any larger size will give the same qualitative results. $\mathrm{MoS}_2$ monolayers are formed of hexagonally packed Mo and S atoms arranged in three layers, the upper and lower ones containing S atoms and the middle one Mo atoms. Monolayers comprise a planar honeycomb lattice with top and bottom S layers arranged in a triangular lattices lying directly upon each other, while the middle Mo layer is also made of a triangular lattice, but rotated by $\pi$. The lattice is shown in Fig.~\ref{fig:2}(b) with big gray dots (molybdenum) and smaller yellow dots (sulphur) forming shifted triangular lattices.
Let us define lattice vectors $R_k$ for the Mo lattice (which are at the same time the nearest neighbors in our tight-binding model): $R_1 = a(1,0)$, $R_2=\frac{a}{2}(1,\sqrt{3})$, $R_3=\frac{a}{2}(-1,\sqrt{3})$ and $R_4=-R_1$, $R_5=-R_2$, $R_6=-R_3$. 

MoS$_2$ monolayers were successfully described by several tight-binding (TB) models of various levels of complexity, with different numbers of orbitals used, including nearest, next-nearest or farther neighbours. Seven\cite{10} or, for better results, eleven\cite{9} (model rederived successfully in \cite{8} and further developed in \cite{16}) orbitals were used for the basis construction in the TB model, to reproduce the low-energy band structure in the entire Brillouin zone, also near the $\Gamma$ point. Although the standard three-band TB model\cite{2} fails around the $\Gamma$ point, it correctly represents the orbital composition around the \textit{K} point near the band edges (both for conduction and valence bands), where Bloch states mainly consist of Mo $d$ orbitals\cite{11,12,13,14,15}. In our case only states near the \textit{K}(\textit{K'}) points will be occupied, thus the three band model is sufficient. This simplest model is successfully employed by other researchers\cite{oskar1,oskar2,3}. 

Consequently, we have described the monolayer structure using only the Mo $d_{z^2}$, $d_{xy}$, $d_{x^2-y^2}$ orbitals and the nearest-neighbors hoppings\cite{2}:
\begin{align}\label{ham1}
H=&\sum_{i\sigma}{\sum^3_{\alpha }{(\epsilon_{\alpha }+{\varphi }_i)}}{\hat{n}}_{i\alpha \sigma }+\nonumber\\
+&\sum_{<ij>\sigma }{\sum_{\alpha \beta }{t_{\alpha \beta }\!\left(\mathrm{\scalebox{0.7}{$R_{k\left(i,j\right)}$}}\right){\hat{c}}^{\dagger }_{i\alpha \sigma }{\hat{c}}_{j\beta \sigma }}}+\nonumber\\
+&\sum_i{\sum_{\sigma \sigma '}{\sum_{\alpha \beta }{{s}_{\sigma \sigma'}{\lambda }_{\alpha \beta }}}}{\hat{c}}^{\dagger }_{i\alpha \sigma }{\hat{c}}_{i\beta \sigma '},
\end{align}
with an assumption $\left({s}_{\sigma\sigma'}\right)\equiv {\sigma }_z$. 
On-site energies $\epsilon_\alpha$ of the $d_{z^2}$ orbital equals $\epsilon_0=1.046$~eV, and of the $d_{xy}$ and $d_{x^2-y^2}$ orbitals equal $\epsilon_1=\epsilon_2=2.104$~eV\cite{2}. 
Additionally, there is an on-site electrostatic potential energy contribution $\varphi_i=\varphi(x_i,y_i)$ in the node $i$ originating from the external gates. The hopping matrix element of an electron from the localized orbital $\beta$ in the $j$-th lattice site to the $\alpha$ localized orbital on the $i$-th unit cell is denoted by $t_{\alpha \beta }\!\left(\mathrm{\scalebox{0.7}{$R_{k\left(i,j\right)}$}}\right)$.
In a compact form we can write (Eq. \ref{ham1}) as:
\begin{equation}\label{ham2}
H=\!\sum_{i}\left(\mathbf{1}_2\otimes D_{i}\!+S\otimes{\varLambda}\right){\hat{n}}_{i}+
\!\!\sum_{<ij>}\mathbf{1}_2\otimes T_{ij}\,{\hat{c}}^{\dagger}_{i}{\hat{c}}_{j}
\end{equation}
On-site on-diagonal matrices $D_i$ and off-diagonal hopping matrices $T_{ij}$ are expressed explicitly as: $(D_{i})_{\alpha\alpha}=\epsilon_{\alpha }+{\varphi }_i$, $(T_{ij})_{\alpha\beta}=t_{\alpha \beta }\!\left(\mathrm{\scalebox{0.7}{$R_{k\left(i,j\right)}$}}\right)$.
The intrinsic spin-orbit interaction\cite{soi1,15,soi2} is represented by
$({\varLambda})_{\alpha\beta}={\lambda }_{\alpha \beta }$, with the spin-orbit coupling parameter $\lambda=0.073$~eV.\cite{2} 
The 3-band model does not reproduce the band crossing (valley inversion) between spin-split states next to the conduction band edge (see e.g. Fig. 3 in [\onlinecite{4}]). 
Thus, to restore the proper ordering of spin-valley states next to the band minimum\cite{23,21}, we put a negative $\lambda = -0.073$~eV in our calculations. 
Let us write an explicit form of $D_i$ and $\varLambda$:
\begin{equation*}
D_i=\begin{bsmallmatrix}
  \varepsilon_{0}+\varphi(x_i,y_i) & 0 & 0 \\
  0 & \varepsilon_1+\varphi(x_i,y_i) & 0 \\
  0 & 0 & \varepsilon_1+\varphi(x_i,y_i)
\end{bsmallmatrix},
\end{equation*}
\begin{equation*}
\varLambda=\begin{bsmallmatrix}
  0 & 0 & 0 \\
  0 & 0 & \imath\lambda \\
  0 & -\imath\lambda & 0
\end{bsmallmatrix},
\end{equation*}
An appropriate Hamiltonian matrix size is given by tensor products $\cdot\otimes\cdot$ with the $z$-Pauli matrix
$\sigma_z\equiv{}S$, $(S)_{\sigma\sigma'}=s_{\sigma\sigma'}$ and the identity matrix
$(\mathbf{1}_2)_{\sigma\sigma'}=\delta_{\sigma\sigma'}$.
This gives an explicit form of the on-site diagonal matrix:
\begin{eqnarray*}
\mathbf{1}_2\otimes D_i\!+S\otimes{\varLambda}\!=\!\!
\begin{bsmallmatrix}
  \varepsilon_{0}+\varphi_i & 0 & 0 & 0 & 0 & 0\\
  0 & \varepsilon_1+\varphi_i & \imath\lambda & 0 & 0 & 0 \\
  0 & -\imath\lambda & \varepsilon_1+\varphi_i & 0 & 0 & 0 \\
  0 & 0 & 0 & \varepsilon_0+\varphi_i & 0 & 0\\
  0 & 0 & 0 & 0 & \varepsilon_1+\varphi_i & -\imath\lambda \\
  0 & 0 & 0 & 0 & \imath\lambda & \varepsilon_1+\varphi_i \\
\end{bsmallmatrix}\!.
\end{eqnarray*}

The hopping elements $t_{\alpha \beta }$ of the off-diagonal $T_{ij}$ matrix depend on the $R_k$ vector, which connects $i$ and $j$ nodes of the lattice. The $R_k$ has six possible directions depending on given $i$ and $j$, numbered with the $k$ index by labeles showed in Fig.~\ref{fig:2}. Thus $k=k(i,j)$ with $k=1\dots6$, and finally $t_{\alpha \beta }=t_{\alpha \beta }\!\left(\mathrm{\scalebox{0.7}{$R_{k\left(i,j\right)}$}}\right)$ what we simply denote by $T=T\!\left(\mathrm{\scalebox{0.8}{$R_{k}$}}\right)$.
Explicitly\cite{2,3}
\begin{equation*}
T\!\left(\mathrm{\scalebox{0.8}{$R_{1}$}}\right)= 
 \begin{bsmallmatrix}
  t_{0} & t_{1} & t_{2} \\
  -t_{1} & t_{11} & t_{12} \\
  t_{2} &-t_{12} & t_{22} \\
\end{bsmallmatrix},
\end{equation*}
\begin{equation*}
T\!\left(\mathrm{\scalebox{0.8}{$R_{2}$}}\right)\!=
\!\!\begin{bsmallmatrix}
  t_{0} & \frac{1}{2}t_{1}+\frac{\sqrt{3}}{2}t_{2} & \frac{\sqrt{3}}{2}t_{1}-\frac{1}{2}t_{2} \\
  -\frac{1}{2}t_{1}+\frac{\sqrt{3}}{2}t_{2} & \frac{1}{4}t_{11}+\frac{3}{4}t_{22} & \hspace{-2mm}\frac{\sqrt{3}}{4}t_{11}-t_{12}-\frac{\sqrt{3}}{4}t_{22} \\
  -\frac{\sqrt{3}}{2}t_{1}-\frac{1}{2}t_{2} & \,\,\,\frac{\sqrt{3}}{4}t_{11}+t_{12}-\frac{\sqrt{3}}{4}t_{22} & \frac{3}{4}t_{11}+\frac{1}{4}t_{22}
\end{bsmallmatrix},
\end{equation*}
\begin{equation*}
T\!\left(\mathrm{\scalebox{0.8}{$R_{3}$}}\right)\!=
\!\!\begin{bsmallmatrix}
  t_{0} & -\frac{1}{2}t_{1}-\frac{\sqrt{3}}{2}t_{2} & \frac{\sqrt{3}}{2}t_{1}-\frac{1}{2}t_{2} \\
  \frac{1}{2}t_{1}-\frac{\sqrt{3}}{2}t_{2} &\frac{1}{4}t_{11}+\frac{3}{4}t_{22} & \hspace{-4mm}-\frac{\sqrt{3}}{4}t_{11}+t_{12}+\frac{\sqrt{3}}{4}t_{22} \\
  -\frac{\sqrt{3}}{2}t_{1}-\frac{1}{2}t_{2} & \,\,-\frac{\sqrt{3}}{4}t_{11}-t_{12}+\frac{\sqrt{3}}{4}t_{22} & \frac{3}{4}t_{11}+\frac{1}{4}t_{22}
\end{bsmallmatrix}\!,
\end{equation*}
\begin{eqnarray}
T\!\left(\mathrm{\scalebox{0.8}{$R_{4}$}}\right)=T\!\left(\mathrm{\scalebox{0.8}{$R_{1}$}}\right)^\intercal\!\!,\quad
T\!\left(\mathrm{\scalebox{0.8}{$R_{5}$}}\right)=T\!\left(\mathrm{\scalebox{0.8}{$R_{2}$}}\right)^\intercal\!\!,\nonumber
\end{eqnarray}
\begin{eqnarray}
T\!\left(\mathrm{\scalebox{0.8}{$R_{6}$}}\right)=T\!\left(\mathrm{\scalebox{0.8}{$R_{3}$}}\right)^\intercal,\nonumber
\end{eqnarray}
with matrix transposition $(.)^\intercal$.
All can be simply derived from the Slater-Koster (two center) interatomic orbitals' elements\cite{7}, with additional nonzero $t_1$ and $t_{12}$ parameters. All parameters are obtained by fitting TB results to those obtained from the first-principes DFT calculations\cite{2}, giving the following hopping integrals: $t_0=-0.184$, $t_1=0.401$, $t_2=0.507$, $t_{11}=0.218$, $t_{12}=0.338$, $t_{22}=0.057$, all in eV.

The employed TB model is sufficient to capture electrons occupying the band-edge in the \textit{K} and \textit{K'} valleys within the Brillouin zone\cite{2}. Such an electron is a quantum information carrier in our nanodevice.

\subsection{Electrostatic Quantum Dot}
The modeled nanodevice is made of a molybdenum disulfide monolayer (deposited on a quartz substrate) covered with a layer of boron nitride insulator on which, subsequently, control gates $\mathrm{G}_{1..4}$ are placed, see Fig.~\ref{fig:1}. For such a structure consisting of various materials, we calculate the electrostatic potential $\phi(\mathbf{r})$ taking into account the voltages applied to the gates $V_{1..4}$ and the highly doped substrate $V_0=0$ together with appropriately chosen boundary conditions. For this purpose we solve the generalized Poisson's equation with the inhomogeneous, space-dependent permittivity $\varepsilon(\mathbf{r})$ of different materials:
\begin{equation}\label{upoiss}
\nabla\cdot\left[\varepsilon_0\varepsilon(\mathbf{r})\nabla\phi_t(\mathbf{r})\right]=-\rho_e(\mathbf{r}).
\end{equation}
From the total potential $\phi_t(\mathbf{r})$, we subtract the electron self-interaction
\begin{equation}\label{upoiss1}
\phi_e(\mathbf{r})=\left(4\pi\varepsilon\varepsilon_0\right)^{-1}\!\int d^2r' \frac{\rho_e(\mathbf{r'})}{|\mathbf{r}-\mathbf{r'}|},
\end{equation}
and thus obtaining $\phi(\mathbf{r})=\phi_t(\mathbf{r})-\phi_e(\mathbf{r})$.
The electron potential energy is $\varphi(\mathbf{r})=-|e|\phi(\mathbf{r})$. For the calculation of the electron eigenstates in the flake we neglect the electron (or hole) charge density putting $\rho_e(\mathbf{r})\equiv 0$. However, later on, during the time evolution, we include the actual time-dependent charge density of the confined electron $\rho_e(\mathbf{r})=-|e|\rho(\mathbf{r},t)$.
We assume the following dielectric constants: $\varepsilon_\mathrm{hBN}=5.1$,\cite{hBN} (for hexagonal boron nitride) and $\varepsilon_{\mathrm{SiO}_2}=3.9$ (for quartz). At the flake level we take the $\varepsilon$ as an average of two neighboring materials: $\varepsilon=(\varepsilon_\mathrm{hBN}+\varepsilon_{\mathrm{SiO}_2})/2=4.5$ ($\varepsilon_0$).\cite{diel}

Appropriately chosen voltages allow for creation of a confinement potential within the monolayer, where a single electron is trapped, forming a QD. Initially we assume voltages $\mathrm{V}_{1..4} =-1500$~mV applied to the gates $\mathrm{G}_{1..4}$ respectively with zero reference voltage on the substrate). By solving the Poisson's equation (\ref{upoiss},\ref{upoiss1}) we obtain an initial potential $\phi(x,y)$ depicted in Fig.~\ref{fig:7} (top left). This way, the energy at $i$-th lattice site includes the potential energy $\varphi$ originating from the external potential $\phi$ at this particular point in the atomic lattice: $\varphi(x_i,y_i)=-|e|\phi(x_i,y_i)$.

The nanostructure also includes two additional electrodes: source \textbf{S} and drain \textbf{D}, depicted in Fig.~\ref{fig:1}, which allow for charge flow through the flake for a given proper polarization. An electron confined in the QD drawn from \textbf{S} charges a capacitor formed by the substrate (i.e. back gate forming the lower plate) and the monolayer (upper plate).

\section{Valley Qubit}
\begin{figure}[b]
	\includegraphics[width=8.7cm]{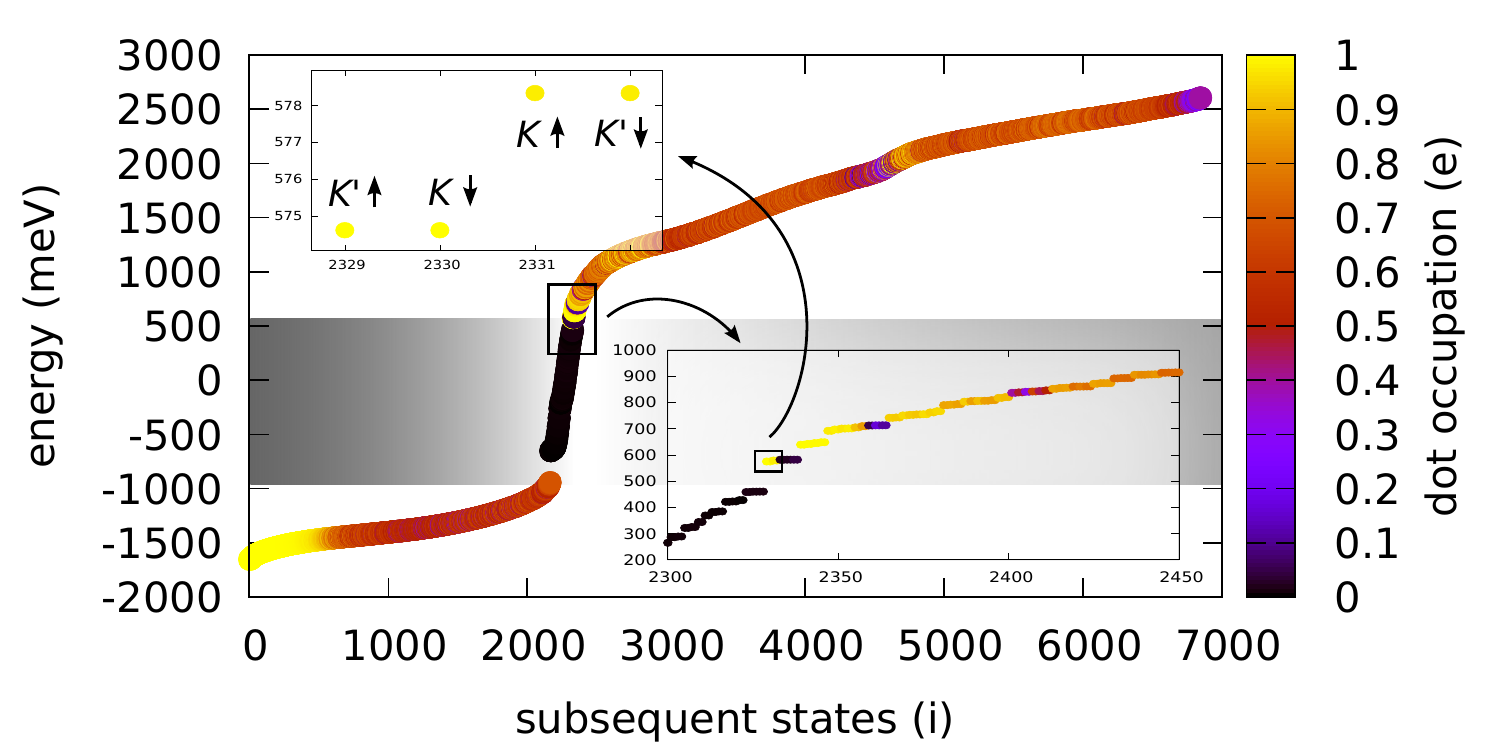}
	\caption{\label{fig:3}
		The electronic eigenvalues for the subsequent states. Colors describe occupation of the confinement (dot) potential localized at the flake center. The black dots denote eigenstates with the electron density on the flake border---edge states, which defines the energy bandgap for electrons (holes) confined within the quantum dot.}
\end{figure}

We diagonalize the Hamiltonian matrix (Eqs. \ref{ham1},\ref{ham2}) for the entire flake lattice obtaining a ladder of subsequent eigenstates $\psi_m$ with their energies, as shown in Fig.~\ref{fig:3}, and electron densities.
The entire hexagonal flake, Fig.~\ref{fig:2} (with the side consisting of $N_B=20$ Mo nodes) is composed of a total number of 1141 nodes, which multiplied by 6 states per node ($N_\alpha=3$ bands $\times$ $N_\sigma=2$ spin-1/2 states) gives a matrix (Eq. \ref{ham2}) size $N_H=1141\times 6=6846$. Diagonalization yields $N_H$ eigenstates.
Additionally, colors are used to mark the QD occupation $\eta_m$ (defined in Eq.~\ref{eta}), namely, the brighter the color is, the more localized the electron is in the center of the flake. Black color marks edge states, that is, states inaccessible to the electron in the QD forming a forbidden energy range, namely, a band gap.


We got a similar state structure as for the triangular flake described in [\onlinecite{3}]. Let us note that there is present a slight space between states near the edge of the valence band. We would get the same space if we merged states for both types of flake edges (circles and squares) in Fig.~5 in [\onlinecite{3}]. This agrees with our results for the hexagonal flake (Fig.~\ref{fig:2}), where both types of edges (i.e. positions of sulphur atoms in relation to molybdenum atoms on the edge) appear alternately, as for the (a) and (b) edge types from Fig.~3 in [\onlinecite{3}].

Knowing $\psi_m$ we construct vectors representing eigenstates $\boldsymbol{\psi}_m(\mathbf{r})=(\psi^{\sigma\alpha}_m(\mathbf{r}))^\intercal$, where $\sigma=1,2$ and $\alpha=1,2,3$, i.e. 
$\boldsymbol{\psi}_m=(\psi^1_m,\psi^2_m,\dots,\psi^6_m)^\intercal$ which belong to the state space ${\cal H}_2^\mathrm{spin}\otimes{\cal H}_3^\mathrm{oribital}$ with spin and three-dimensional Mo orbitals space.

To identify states we need an electron density both in the position space $\rho(x,y)$ and in the reciprocal space $\tilde{\rho}(k_x,k_y)$. Information about spin in a particular state is also helpful for full identification. The electron density for a particular spin and orbital $|\psi^{\sigma\alpha}_m(\mathbf{r})|^2$ allows to calculate a density for the $m$-th eigenstate as:
\begin{equation}\label{rho}
\rho_m(\mathbf{r})=\sum_{\sigma\alpha}|\psi^{\sigma\alpha}_m(\mathbf{r})|^2,
\end{equation}
where $\sigma=1,2$ is a spin index and $\alpha=1,2,3$ is a Mo-orbital index, $\mathbf{r}\equiv(x,y)$. Now we can define an occupancy coefficient for the QD. We assume the QD radius $r_\mathrm{QD}=2.3$~nm,
\begin{equation}\label{eta}
\eta_m=\int_0^{r<2r_\mathrm{QD}}\!d^2r\,\rho_m(\mathbf{r}),
\end{equation}
where $r=\sqrt{x^2+y^2}$. We select the coefficient $2$ to color all the states which are not edge states, while the latter ones are marked as black. 

Spin component of the $m$-eigenstate is defined as:
\begin{align}\label{spin}
\langle s_i \rangle_m &= \frac{\hbar}{2}\langle\boldsymbol{\psi}_m|\sigma_i\otimes\mathbf{1}_3|\boldsymbol{\psi}_m\rangle = \nonumber\\
&= \int_F d^2r\,\boldsymbol{\psi}^\dagger_m(\mathbf{r})\,\sigma_i\otimes\mathbf{1}_3\,\boldsymbol{\psi}_m(\mathbf{r}).
\end{align}
with the Pauli matrices $\sigma_i$, and flake surface $F$. 

For selected stationary states we calculate the Fourier transform:
\begin{equation}\label{fst}
\Phi_m(\mathbf{k})=\int_F d^2r\,\boldsymbol{\psi}_m(\mathbf{r})\,e^{-\imath \mathbf{k}\mathbf{r}},
\end{equation}
where $\mathbf{k}\equiv(k_x,k_y)$, $\Phi_m(\boldsymbol{k})=(\Phi^{\sigma\alpha}_m(\boldsymbol{k}))^\intercal$ and the k-area in the reciprocal space $\tilde{F}:k_{x,y}\in[-\frac{2\pi}{a},\frac{2\pi}{a}]$. Now the density in the reciprocal space is expressed as $\tilde{\rho}_m(\mathbf{k})=|\Phi_m(\mathbf{k})|^2=\sum_{\sigma\alpha}|\Phi_m^{\sigma\alpha}(\mathbf{k})|^2$. 
Take a note that the integral is taken over a finite flake (or equivalently over the entire space but with vanishing electron density beyond the flake).
\begin{figure}[h]
	\includegraphics[width=4.1cm]{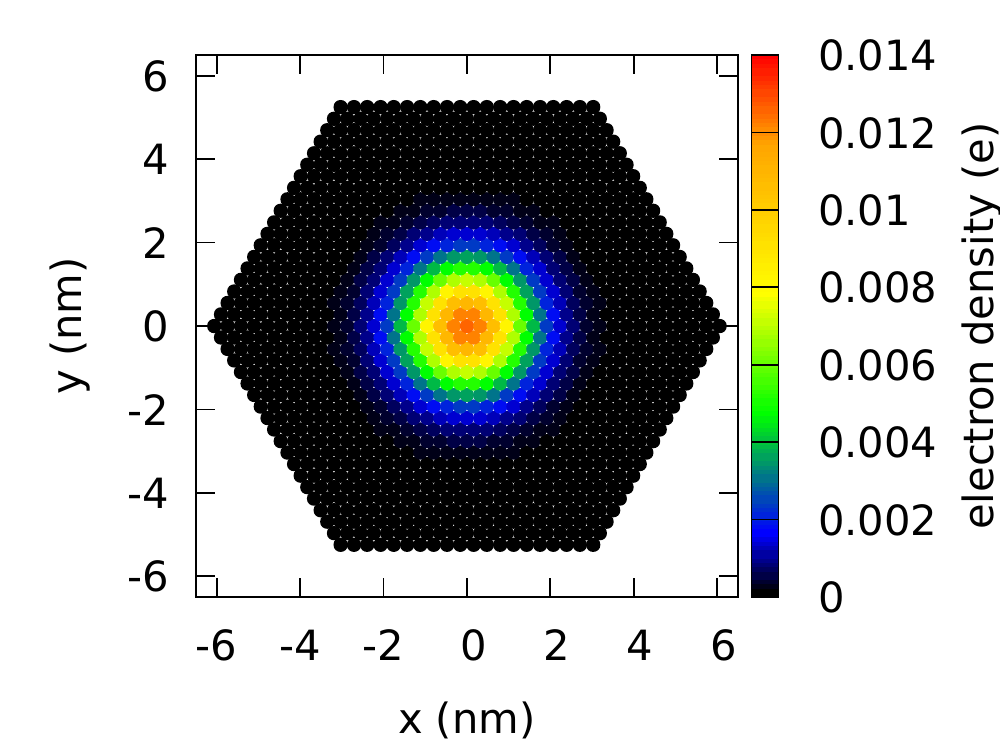}
	\includegraphics[width=3.5cm]{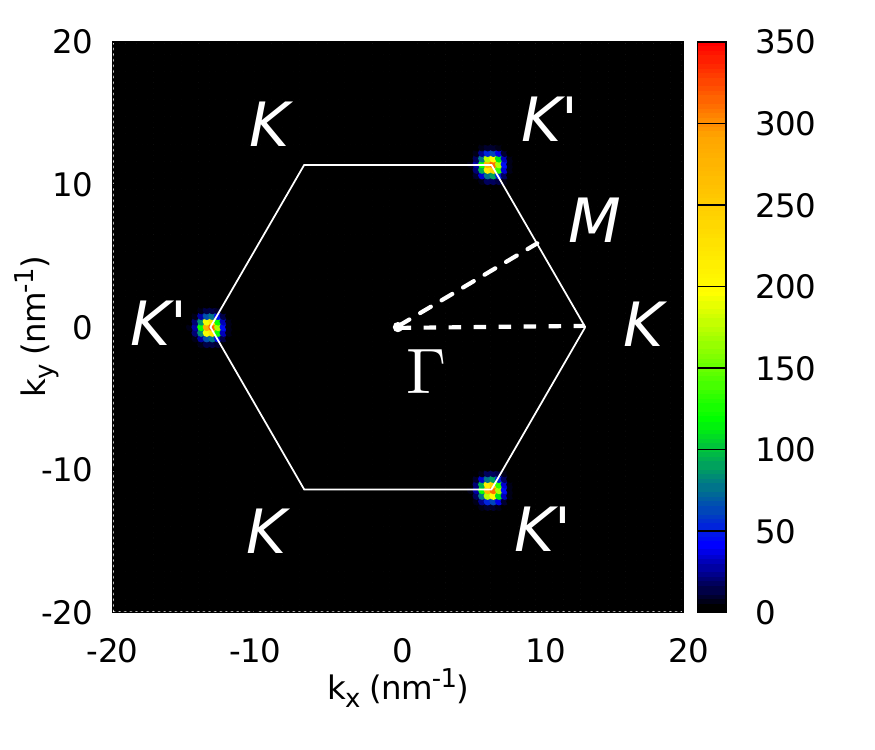}\\
	\includegraphics[width=2.4cm]{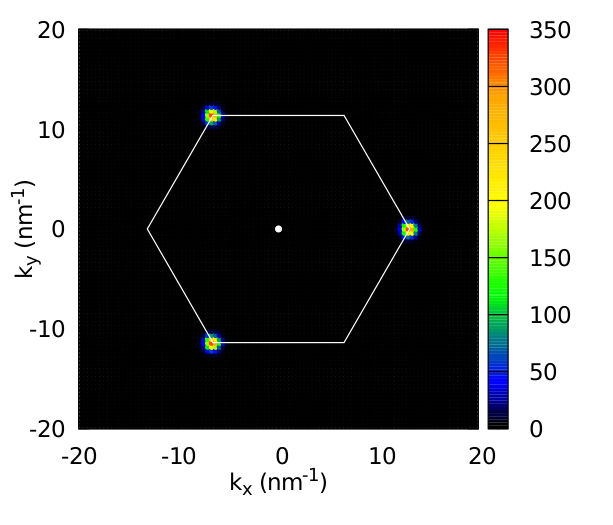}
	\includegraphics[width=2.4cm]{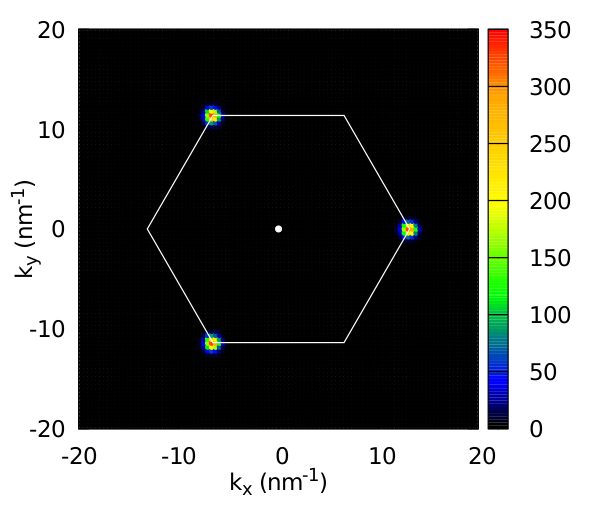}
	\includegraphics[width=2.4cm]{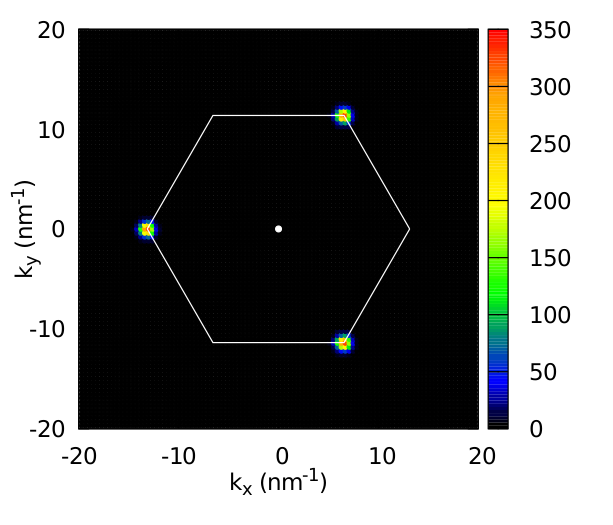}
	\caption{\label{fig:4} First four electron states in the conduction band, marked with dots in the upper inset from Fig.~\ref{fig:3}. The electron density $\rho$ (top left) is the same as for the remaining four. The density $\tilde{\rho}$ in the reciprocal space for the first state  is located at \textit{K'} (with spin-up) (top right with marked Brillouin zone with high symmetry points). For the second state (bottom left) it is located at \textit{K} point (spin-down); and vice versa for the second pair: \textit{K} (spin-up) and \textit{K'} (spin-down).}
\end{figure}

Now let us take a look at the first four states from the conduction band (for $m =2329$, 2330, 2331, 2332) marked in the close-ups in the lower and upper small insets in Fig.~\ref{fig:3}. Their electron densities (obtained with Eq.~\ref{rho}) $\rho(\mathbf{r})$ are shown in Fig.~\ref{fig:4} (top left). If we do the Fourier transform (Eq. \ref{fst}) of the wave function of each state we obtain $\tilde{\rho}(\mathbf{k})$ in the Brillouin zone (BZ). Also, we determine their spins (Eq. \ref{spin}). It turns out that the density of the first state (top right) is localized at the \textit{K'} point (with spin up) in the BZ, while the next state is located at \textit{K} (spin down). Next pair of states separated by SOI splitting are localized at \textit{K} (spin up) and \textit{K'} (spin down). We use two spin-up states from \textit{K} and \textit{K'} to define a \textit{valley qubit}. The width of the SOI splitting between 1st, 2nd and 3rd, 4th states equals here $2\Delta\sim3.7$~meV, and is reproduced correctly\cite{15}. We also mark the BZ in Fig.~\ref{fig:4}, along with points of high symmetry: $\Gamma$ in the center of the zone, two types of \textit{K}(\textit{K'}) at the corners of the hexagonal zone and $M$ on the edges of the hexagon. The coordinates of the high symmetry points: $\Gamma=(0,0)$, one of $K=\frac{\pi}{a}(\frac{4}{3},0)$ and one of $M=\frac{\pi}{a}(1,\frac{1}{\sqrt{3}})$, with the lattice constant $a=0.319$~nm.

\begin{figure}[h]
	\includegraphics[width=2.5cm]{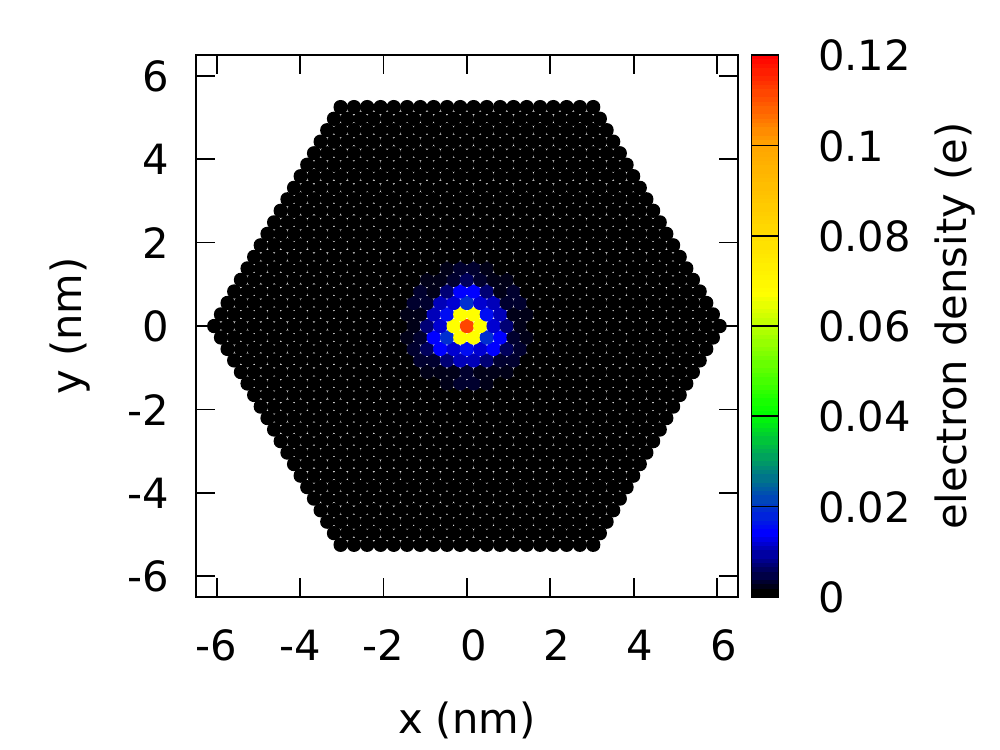}
	\includegraphics[width=2.5cm]{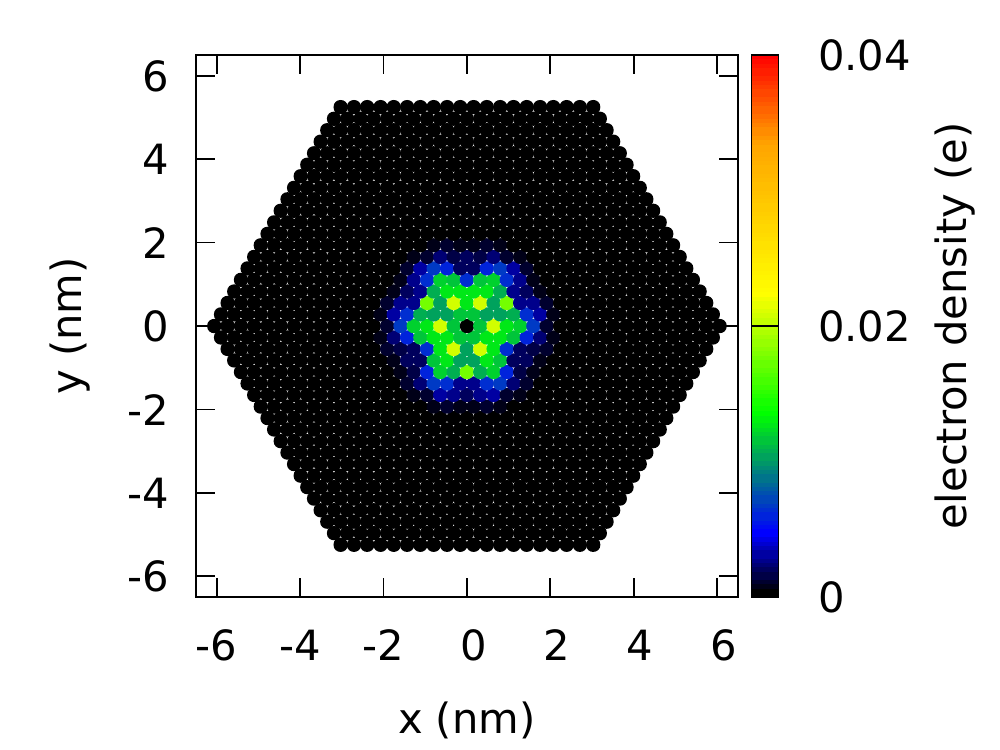}
	\includegraphics[width=2.5cm]{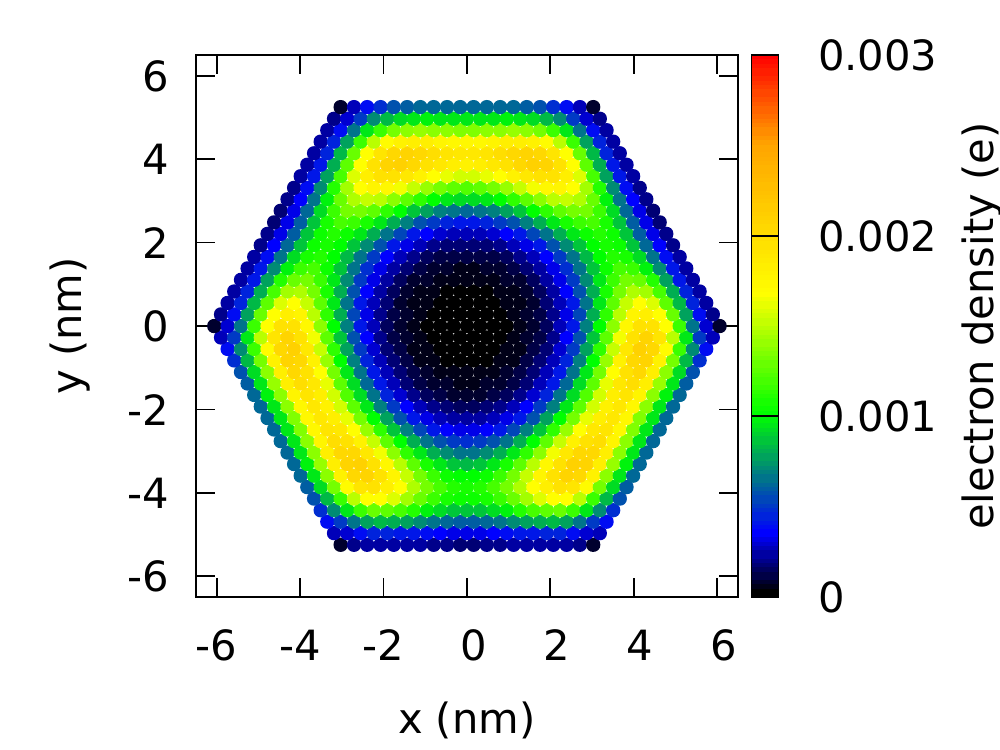}\\
	\includegraphics[width=2.5cm]{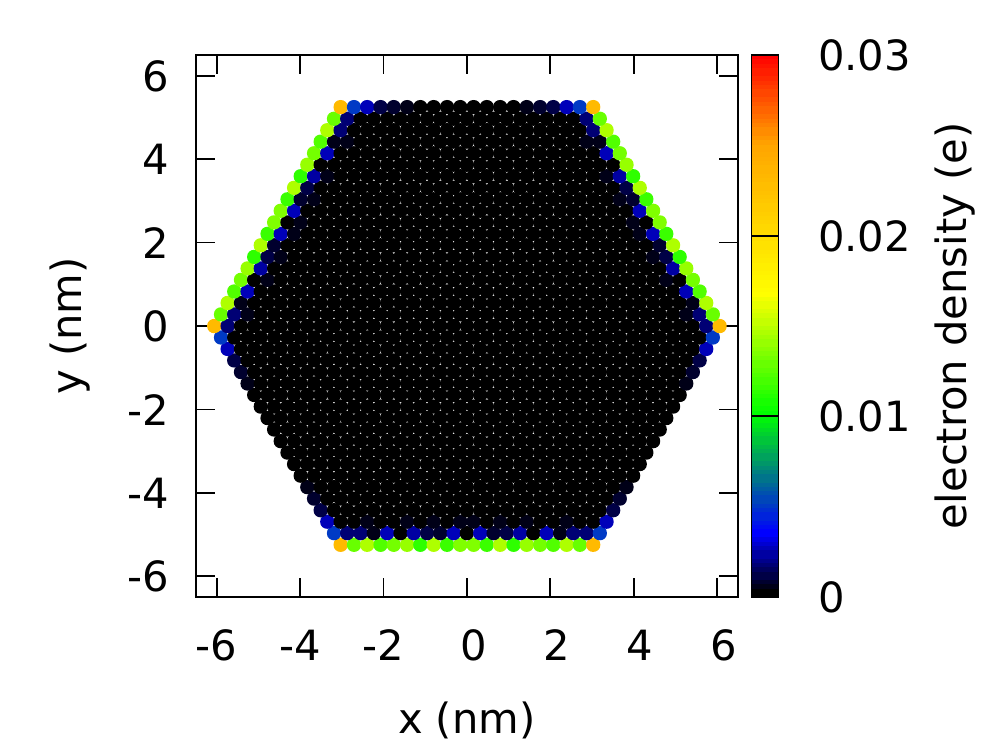}
	\includegraphics[width=2.5cm]{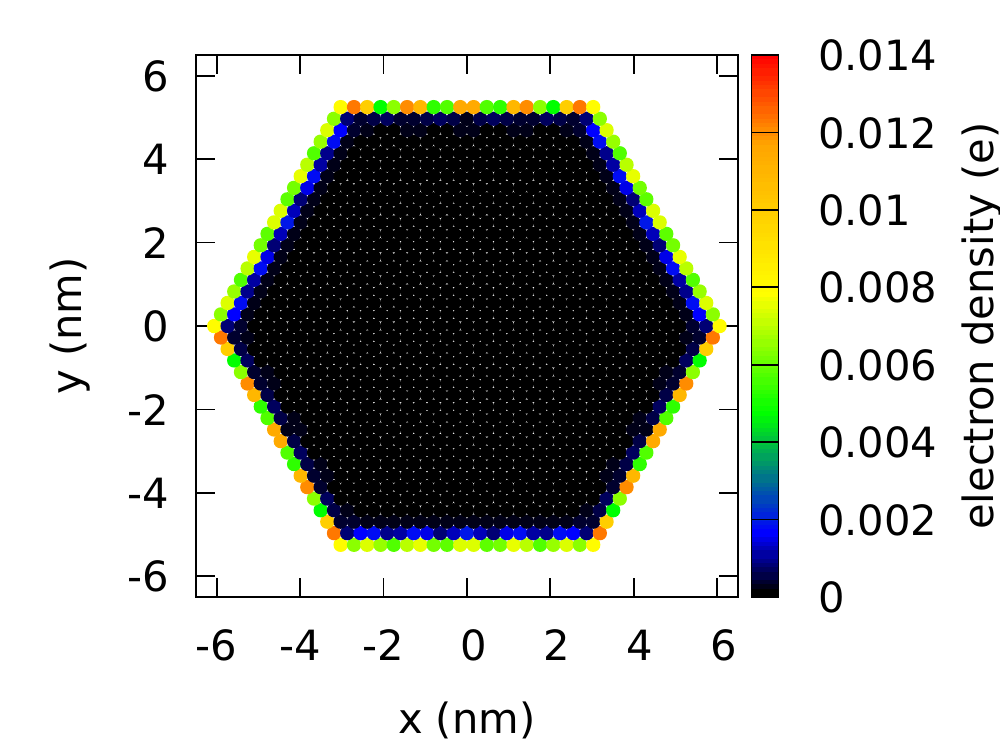}
	\includegraphics[width=2.5cm]{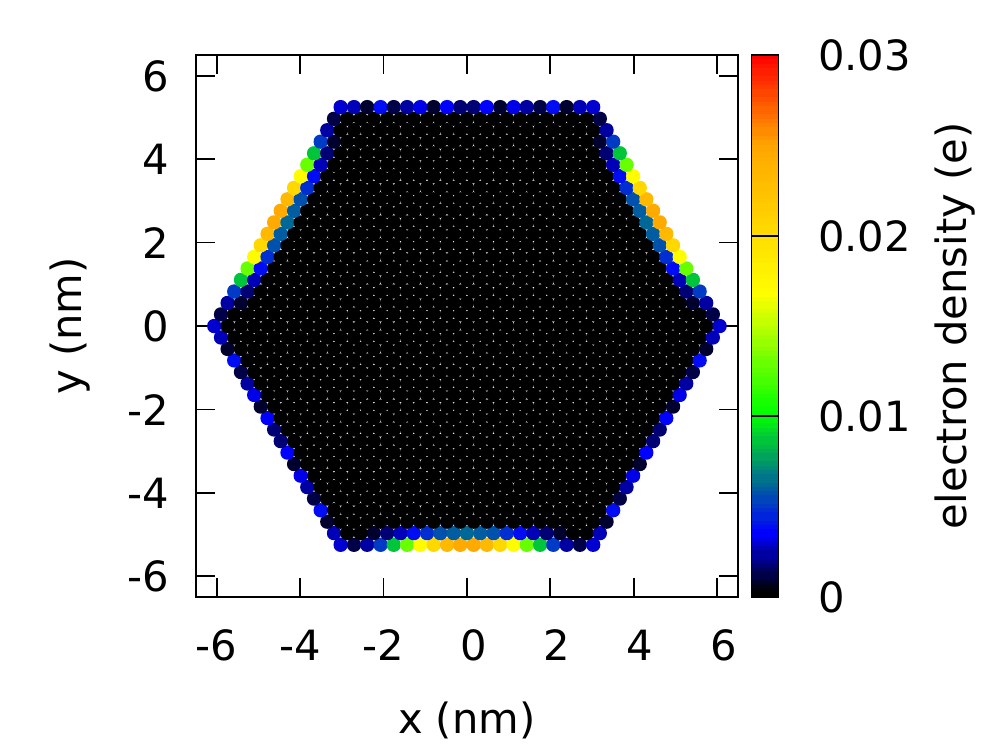}\\
	\includegraphics[width=2.5cm]{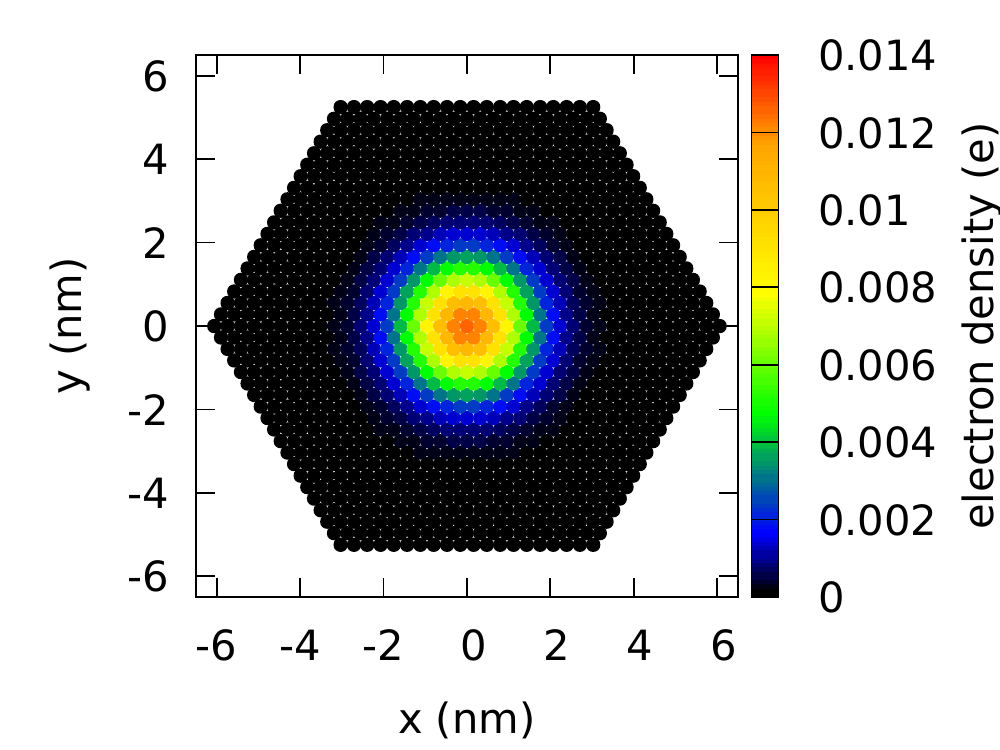}
	\includegraphics[width=2.5cm]{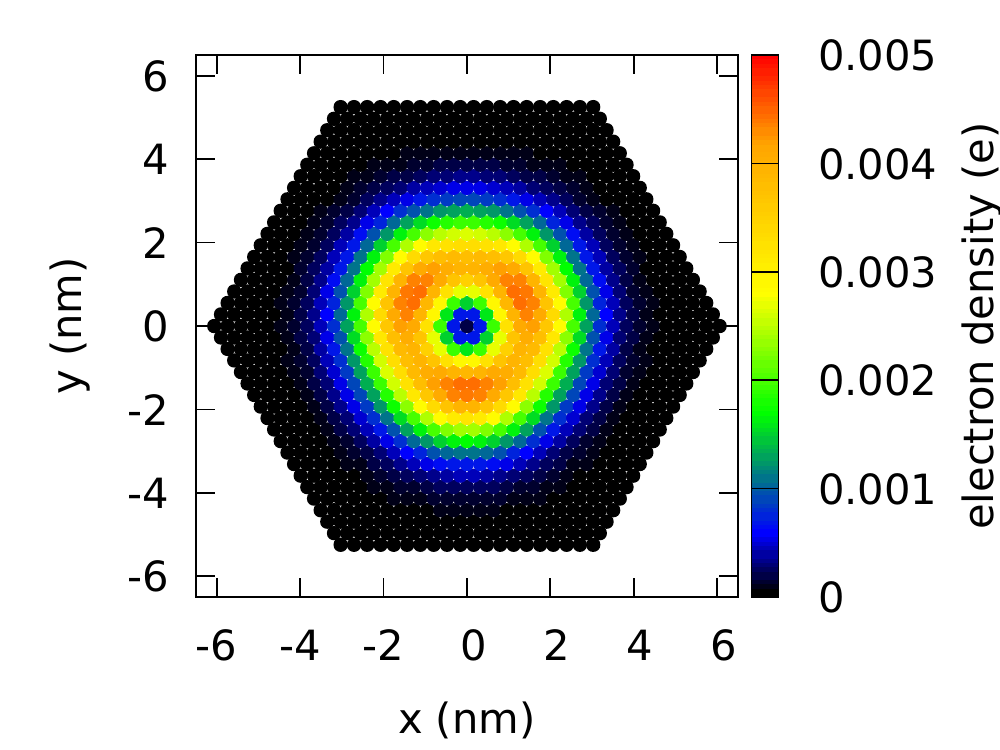}
	\includegraphics[width=2.5cm]{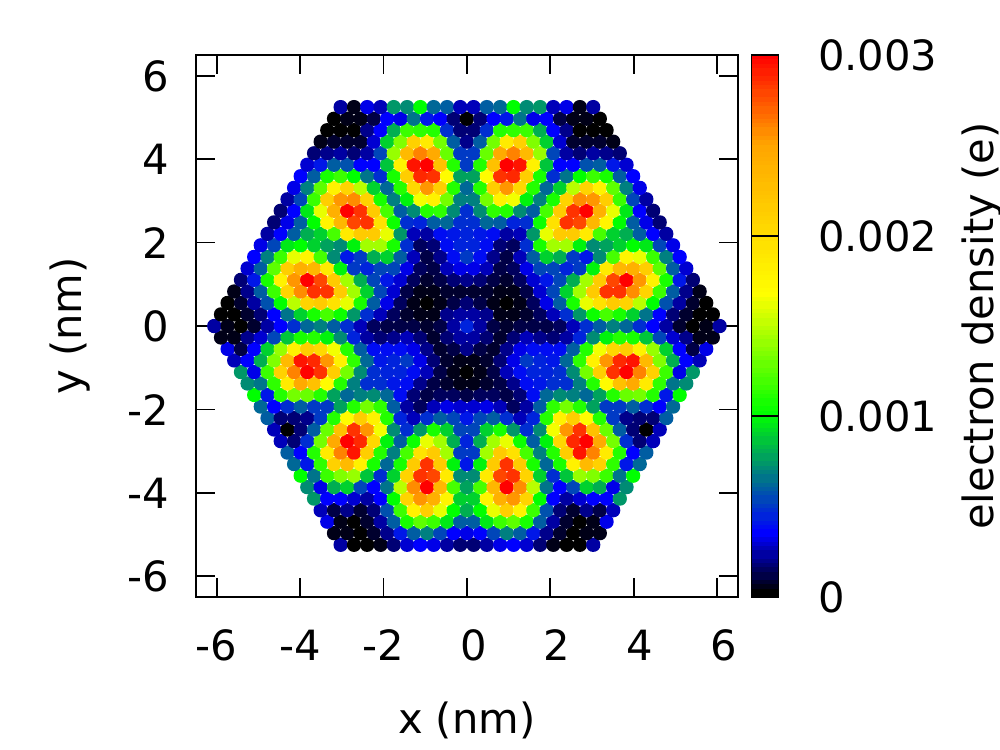}\\
	\includegraphics[width=2.5cm]{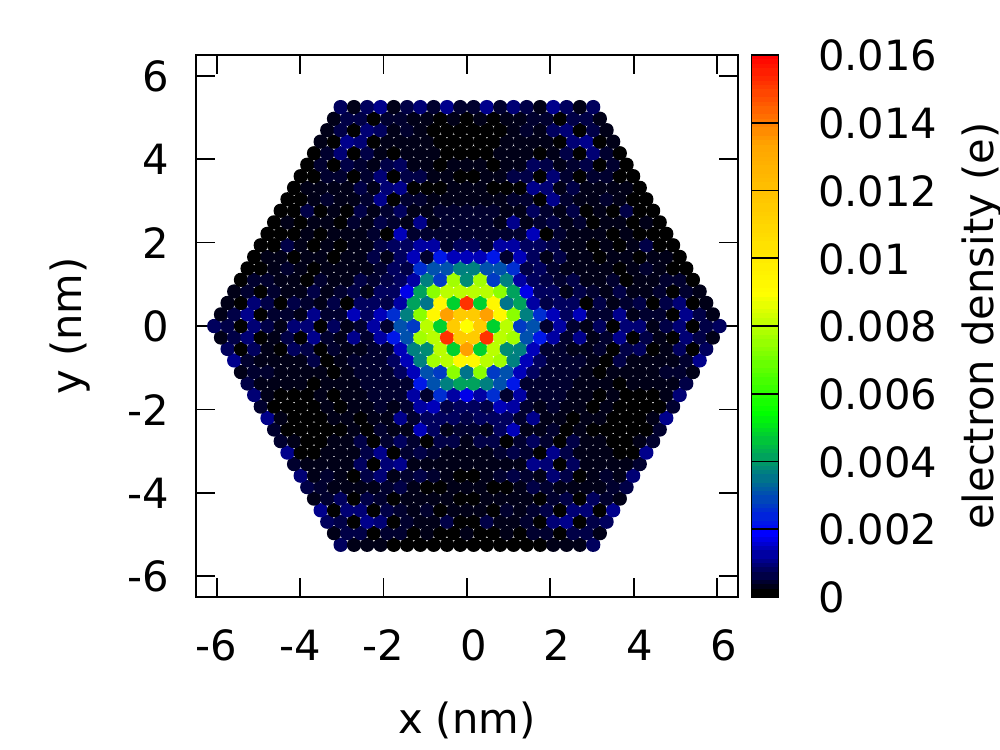}
	\includegraphics[width=2.5cm]{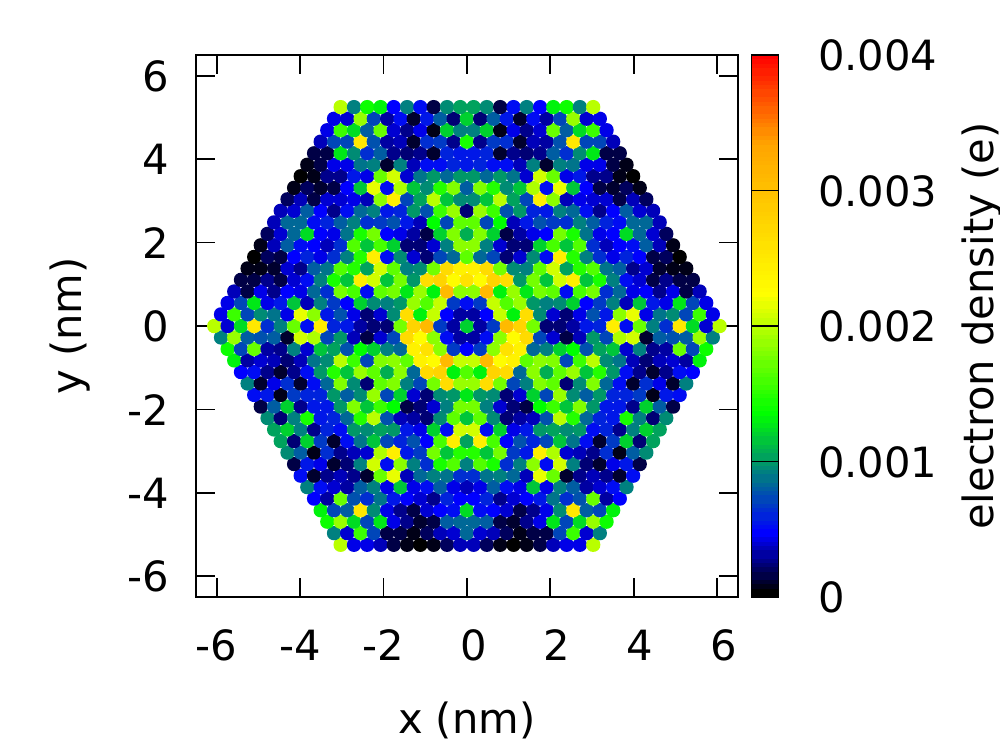}
	\includegraphics[width=2.5cm]{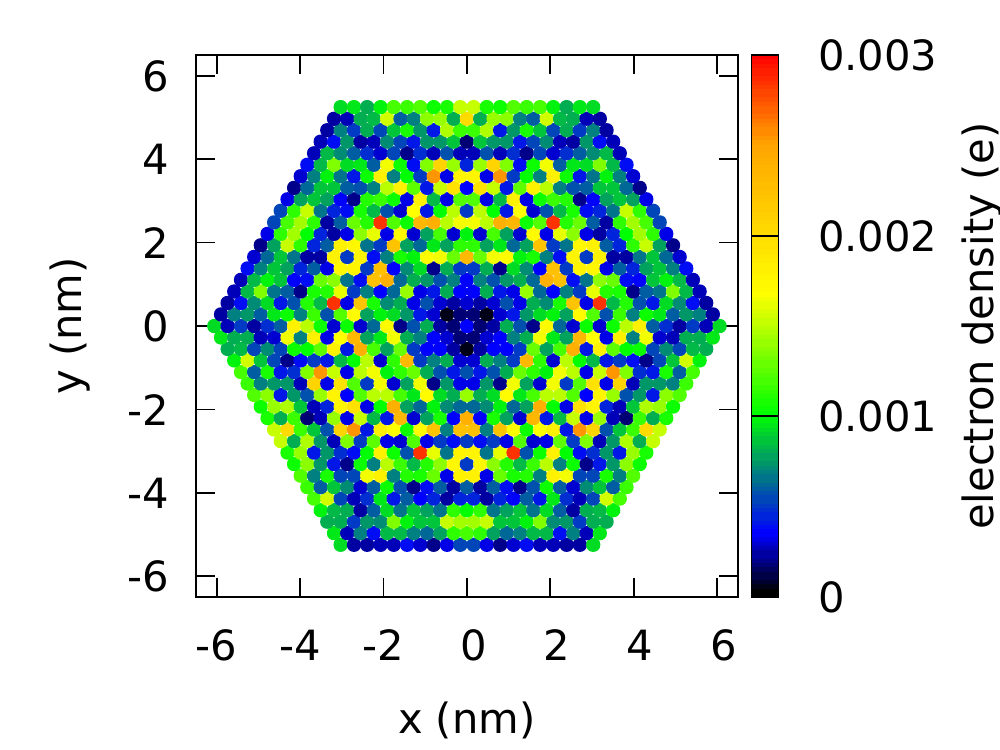}\\
	\caption{\label{fig:5} The densities $\rho(x,y)$ of selected eigenstates from Fig.~\ref{fig:3}, for the valence band (first row with numbers: $m=2,33,2165$), edge states (second row: 2237, 2277, 2318) and conduction band (3rd and 4th one: 2330, 2340, 2450, 2570, 2600, 3200).}
\end{figure}

\begin{figure}[h]
	\includegraphics[width=2.4cm]{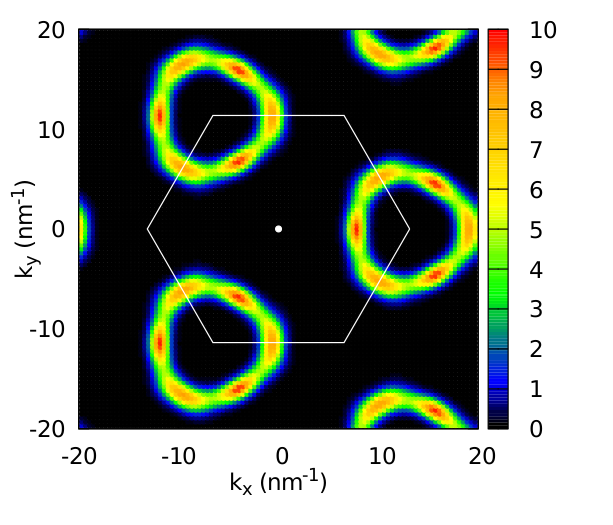}
	\includegraphics[width=2.4cm]{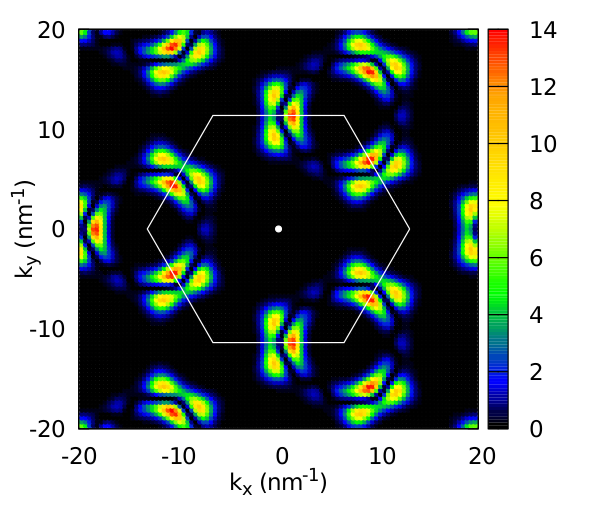}
	\includegraphics[width=2.4cm]{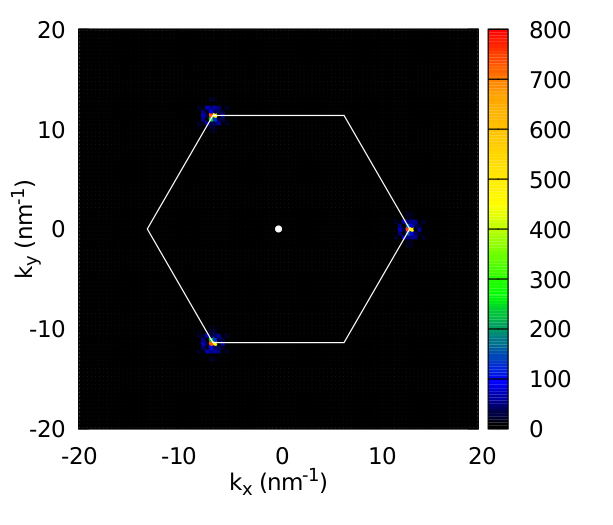}\\
	\includegraphics[width=2.4cm]{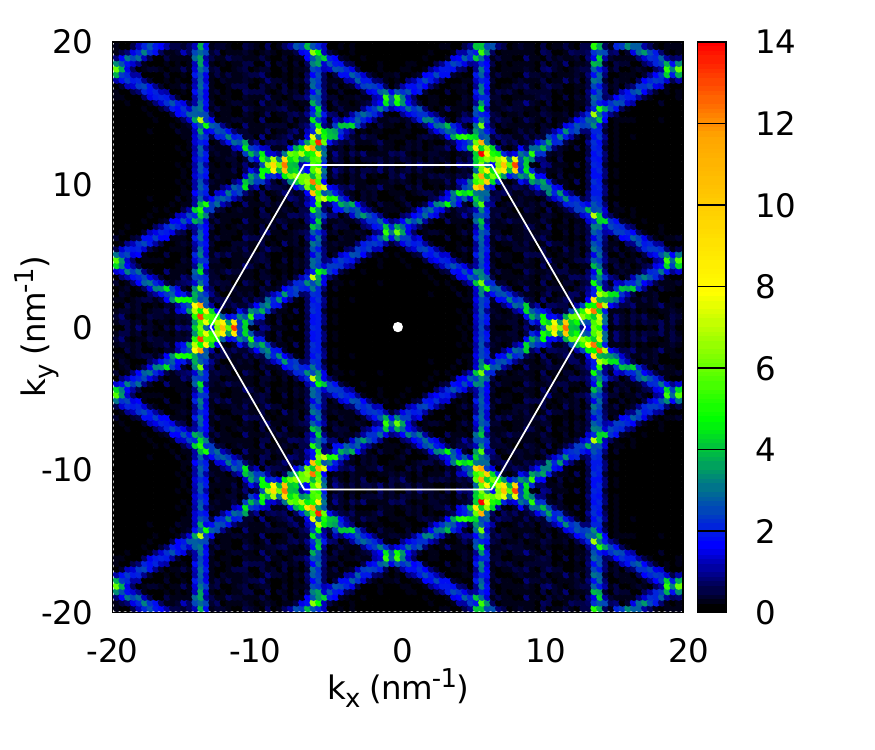}
	\includegraphics[width=2.4cm]{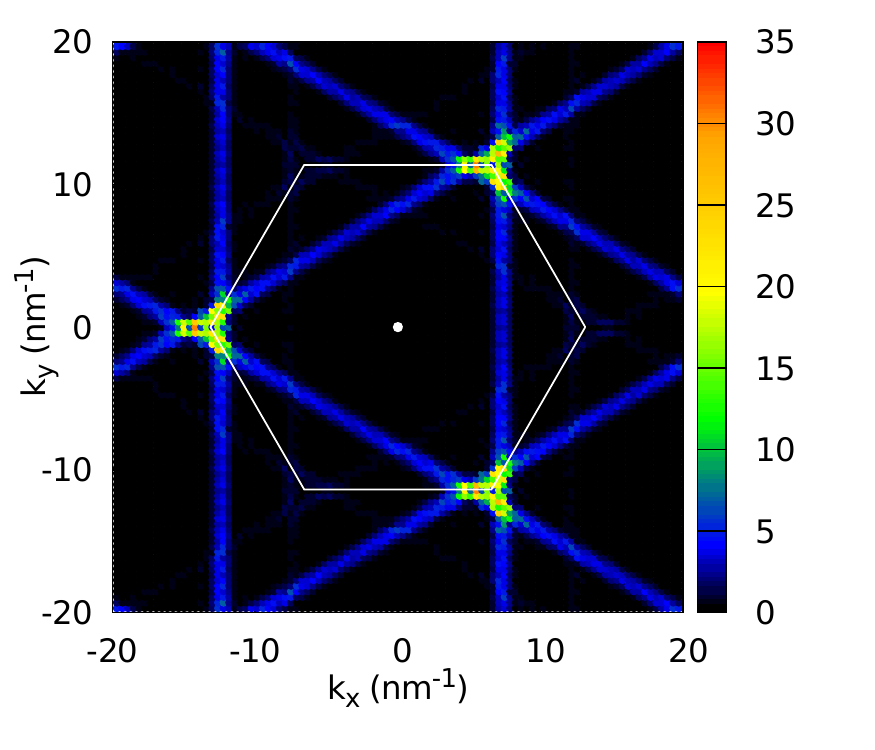}
	\includegraphics[width=2.4cm]{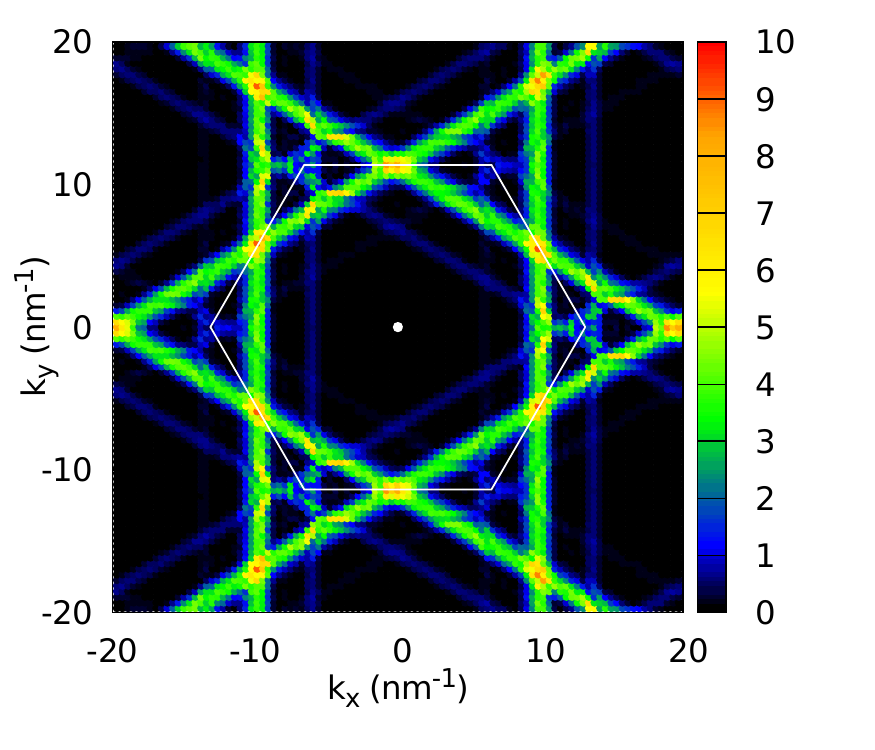}\\
	\includegraphics[width=2.4cm]{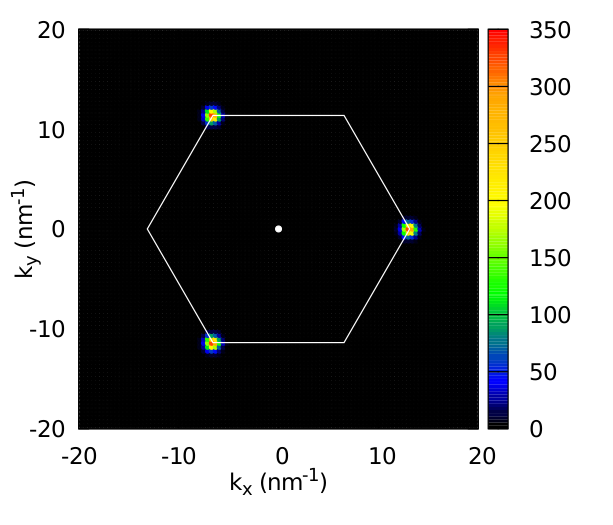}
	\includegraphics[width=2.4cm]{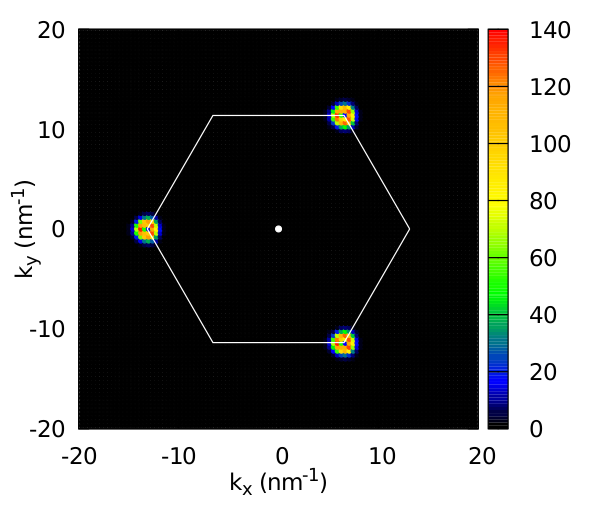}
	\includegraphics[width=2.4cm]{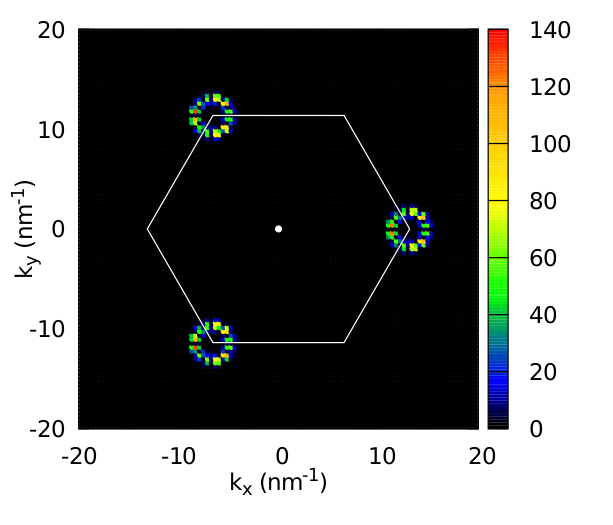}\\
	\includegraphics[width=2.4cm]{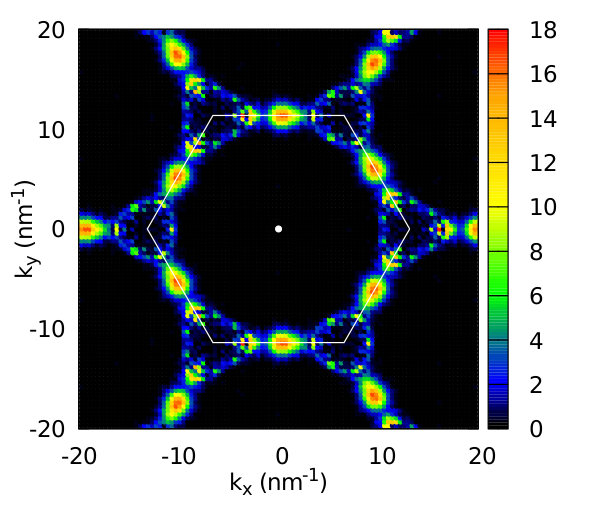}
	\includegraphics[width=2.4cm]{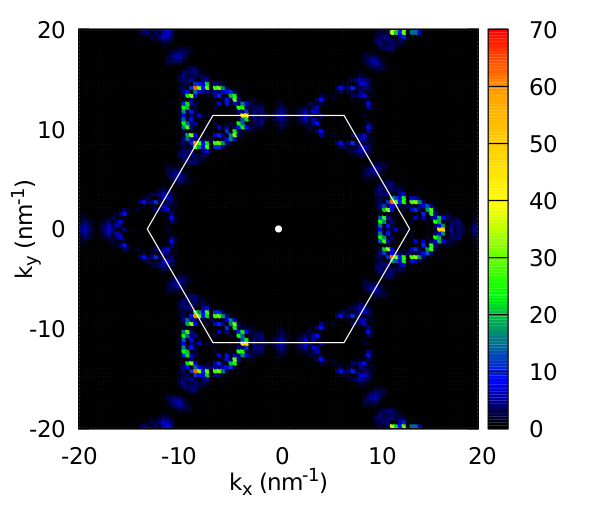}
	\includegraphics[width=2.4cm]{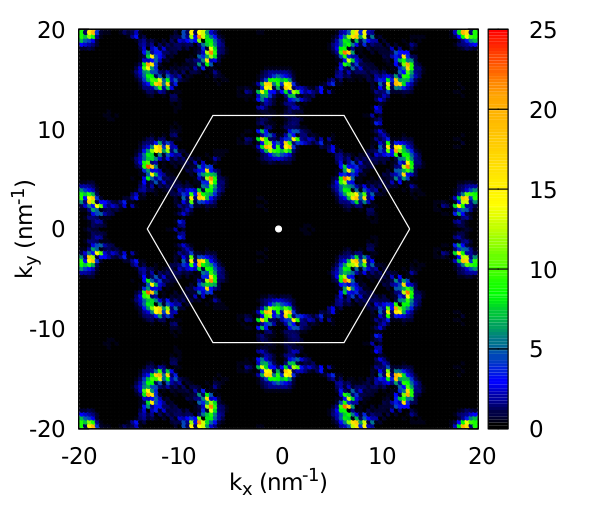}\\
	\caption{\label{fig:6} Corresponding densities of the Fourier transforms $\tilde{\rho}(k_x,k_y)$ of the selected eigenstates (like in Fig.~\ref{fig:5}) in the reciprocal space. The order of subsequent densities is the same as in Fig.~\ref{fig:5}.}
\end{figure}

Let us focus on the remaining eigenstates for a bit. In Fig.~\ref{fig:5} we see densities in the position space $\rho(\mathbf{r})$ and in Fig.~\ref{fig:6} their corresponding densities in the reciprocal space $\tilde{\rho}(\mathbf{k})$ for various eigenstates. In the top row, hole states: 1 and 2 are located deep in the valence band with densities localized around the point $M$ and on the line $K-\Gamma$  (Fig.~\ref{fig:6}), while 3 is a state strongly repulsed toward the edges by the (repulsive for holes) confinement potential (Fig.~\ref{fig:5}) located near the top of the valence band in the valley \textit{K}. In the second row we see the edge states localized at the edge of the flake, characterized by quite complex densities $\tilde{\rho}$ with various symmetries. These are trivial edge states which are briefly explained in the appendix to this paper. This means that MoS$_2$ is a trivial (non-topological) insulator. In the third and fourth rows we have electron states of energies gradually moving away from the edge of the conduction band. In the third row $\tilde{\rho}$ drifts away from the \textit{K}(\textit{K'}) point, while in the fourth, for higher energies, the densities nearly reach the point $M$ in the BZ.

Two inequivalent points \textit{K} and \textit{K'} at the corners of the BZ are related to two inequivalent triangular Bravais lattices. However the M points are equivalent, as any of them can be obtained from any other using vectors of the reciprocal space. It is a rule that for states around the \textit{K}(\textit{K'}) points we have four states of different valleys and spins but of the same densities $\rho$. On the other hand, near the $M$ point, we have a pair of states for each same $\rho$ (only spin degeneracy).


An addition of the valley degree of freedom opens a multitude of possibilities to define a qubit in similar nanostructures. First and foremost we have spin and the spin qubit\cite{var1}, then we also have a qubit defined using the valley degree of freedom\cite{17}. It turns out that we can make a hybrid spin-valley qubit\cite{4}, for which one basis state is a state in the valley \textit{K'} and with spin up, while the second state is one in the valley \textit{K} and with spin down. These are degenerate states, however they can be separated in a controllable manner by applying an appropriate magnetic field.

However, it is of utmost importance to be able to create multi-qubit registers and also perform two-qubit operations on them. Any multi-qubit operation can be approximated with single and two qubit operations.\cite{barenco} In a system consisting of multiple hybrid qubits we can realize two-qubit swap gate by exchanging spins between adjacent electrons. 
Nevertheless, it appears that the most interesting and generic two qubit system is one which exploits both spin and valley degrees of freedom of a charge carrier, either an electron or a hole, spanning a two qubit system of such four basis states\cite{burk,var3}.

\section{Single Qubit Gate}
In previous sections we defined a valley qubit, carried by an electron trapped in an electrostatic quantum dot, initially localized in the \textit{K} valley within the Brillouin zone (see Fig.~\ref{fig:4}). We now investigate the time evolution of such a system and show that such a nanostructure can be used to perform logic operations in a controllable manner. We will perform the NOT operation on our qubit, and halfway through the NOT gate we will also get the Hadamard operation. It is achieved all electrically solely using oscillatory voltages applied to the local gates and this provides a reliable means to control the state of the qubit.

\subsection{Calculation of Time Evolution}
A time-dependent wave function, for a particular spin $\sigma$ and orbital $\alpha$ can be expressed as a linear combination of basis states with time-dependent amplitudes $c_n(t)$ and phase factors: 
\begin{equation}\label{psit}
\Psi^{\sigma\alpha}(\mathbf{r},t)=\sum_{n}c_{n}(t)\,\psi^{\sigma\alpha}_{n}(\mathbf{r})\,e^{-\frac{\imath}{\hbar}E_{n}t},
\end{equation}
where $n=1,\dots,N_\mathrm{base}$. We assume a basis consisting of $N_\mathrm{base}=200$ first eigenstates in the conduction band. Increasing this number does not change the results, but keeping it small allows for faster calculations. Using the wave function (Eq. \ref{psit}) we calculate a time-dependent electron density:
\begin{equation}\label{rot}
\rho(\mathbf{r},t)=\sum_{\sigma\alpha}|\Psi^{\sigma\alpha}(\mathbf{r},t)|^2.
\end{equation}
The full spinor vector has the form:
$\boldsymbol{\Psi}(\mathbf{r},t)=(\Psi^{\sigma\alpha}(\mathbf{r},t))^\intercal$.
Time evolution is governed by the time-dependent Schr\"odinger equation:
\begin{align}\label{fulls}
\imath\hbar\frac{\partial}{\partial t}\boldsymbol{\Psi}(\mathbf{r},t)=H'(\mathbf{r},t)\boldsymbol{\Psi}(\mathbf{r},t),
\end{align}
with time-dependent Hamiltonian being a sum of a stationary Hamiltonian (Eq. \ref{ham1}), with the previously found eigenstates $\boldsymbol{\psi}_m$: 
$H(\mathbf{r})\boldsymbol{\psi}_m(\mathbf{r})=E_m \boldsymbol{\psi}_m(\mathbf{r})$, and a time-dependent potential energy contribution $\delta\varphi$:
\begin{align}
H'(\mathbf{r},t)=H(\mathbf{r})+\delta\varphi(\mathbf{r},t).
\end{align}
That is, the full time-dependent potential energy
\begin{align}\label{fullp}
\varphi'(\mathbf{r},t)=\varphi(\mathbf{r})+\delta\varphi(\mathbf{r},t),
\end{align} 
contains an oscillating part $\delta\varphi(\mathbf{r},t)$, which is generated by modulation of the gate voltages. Let us remind that the potential energy is calculated as $\varphi'(\mathbf{r},t)=-|e|\phi(\mathbf{r},t)$. During the evolution of $\boldsymbol{\Psi}(\mathbf{r},t)$ according to the presented Schr\"odinger equation (\ref{fulls}), the potential $\phi(\mathbf{r},t)$ is obtained within the mean field approximation \cite{kot}. It is done by solving the generalized Poisson's equation (\ref{upoiss},\ref{upoiss1}) for $\rho(\mathbf{r},t)$ at every time step. The potential includes the induced charge and thus takes into account the self-focusing effect. On the other hand $\rho(\mathbf{r},t)$ depends on the actual shape of the wave function at a particular moment, hence the Poisson's and Schr\"odinger's equations must be solved in a self-consistent way. To obtain the stationary eigenstates we take into account only the confinement potential $\phi(\mathbf{r})$ originating from applied voltages. We neglect the electron (hole) charge. That is why eigenstates $\boldsymbol{\psi}_m(\mathbf{r})$ do not account for electron self-focusing (through metallic gates).

Now, knowing $\varphi'(\mathbf{r},t)$ and $\varphi(\mathbf{r})$ we can employ (Eq. \ref{fullp}) to calculate $\delta\varphi(\mathbf{r},t)$. Putting (Eq. \ref{psit}) into (Eq. \ref{fulls}) we get a formula allowing to calculate expansion coefficients at subsequent moments of time:
\begin{align}\label{ct}
c_m^{\,t+1}\!=c_m^{\,t-1}\!-2\frac{\imath\Delta t}{\hbar}\sum_n c_n^{\,t}\, \langle \boldsymbol{\psi}_m| \delta\varphi(\mathbf{r},t)|\boldsymbol{\psi}_n \rangle\, e^{-\frac{\imath}{\hbar}(E_{n}\!-E_m)t},
\end{align}
where $c_m(t)\equiv c^{\,t}_m$. We used here a central time derivative approximation $\frac{\partial}{\partial t}c_m(t)\simeq \frac{c^{t+1}_m-c^{t-1}_m}{2\Delta t}$.

Hamiltonian for each time step includes the time-dependent difference of the confinement potential $\delta\varphi(\mathbf{r},t)$. The matrix elements for this difference  $\delta\varphi_{mn}(t)=\langle \boldsymbol{\psi}_m| \delta\varphi(\mathbf{r},t)|\boldsymbol{\psi}_n \rangle$ 
are calculated at every time step, to solve the next time step by using (Eq. \ref{ct}).
This approach, however, is time consuming and in practice we can proceed in a different way.
At first we calculate the matrix elements at every time step of the evolution. These calculations are done only for a single cycle $T$ of voltage changes, i.e. $t\leq T$.\cite{note1}
What is important, we also have to remember all the matrix elements $\delta\varphi_{mn}(t_i)\equiv\delta\varphi_{mn}^i$ at equidistant time periods $t_i=i\frac{T}{N_\varphi}$, $i= 0..N_\varphi$. Now, for the rest of the evolution (i.e. for subsequent cycles) we calculate $\delta\varphi_{mn}$ smoothly interpolating the values between subsequent time moments $t_i$ and $t_{i+1}$:
\begin{align}
2\delta\varphi_{mn}(t) =\,\,&\delta\varphi_{mn}^i+\delta\varphi_{mn}^{i+1}\,+ \nonumber\\
+ &\left(\delta\varphi_{mn}^i-\delta\varphi_{mn}^{i+1}\right)\cos\!\left(\frac{t-t_i}{T}N_\varphi\pi\right),
\end{align}
for $t\in[t_i,t_{i+1}]$.
Thanks to smooth transition $N_\varphi$ does not have to be large (we assumed $N_\varphi=16$). This approximation is valid as long as the voltage pumping process is adiabatic and the electron does not gain energy during the evolution.

The time-dependent Schr\"odinger equation is solved with a predictor-corrector method, where the predictor is calculated using the Richardson (leapfrog) scheme, aka Askar-Cakmak method\cite{28}; and the corrector is given by the implicit Crank-Nicolson method.

The Fourier transform is also calculated for the time-dependent wave function, in a similar way as in (Eq. \ref{fst}):
\begin{equation}\label{fout}
\Phi^{\sigma\alpha}(\mathbf{k},t)=\int_F d^2r\,\Psi^{\sigma\alpha}(\mathbf{r},t)\,e^{-\imath \mathbf{k}\mathbf{r}},
\end{equation}
where $\mathbf{k}\equiv(k_x,k_y)$, and k-area $\tilde{F}:k_{x,y}\in[-\frac{2\pi}{a},\frac{2\pi}{a}]$. Now the charge density in the reciprocal space has the form $\tilde{\rho}(\mathbf{k},t)=\sum_{\sigma\alpha}|\Phi^{\sigma\alpha}(\mathbf{k},t)|^2$. 
We are ready to calculate a time-dependent valley index. Let us define an angle $\vartheta=\mathrm{atan2}(k_y,k_x)$. To calculate the index we need only take into account 1/3 of the reciprocal space area $\tilde{F}$ (two opposite $\pi/3$ sectors, encompassing exactly one \textit{K} point and one \textit{K'} point), i.e. for  
$\tilde{F}_{1/3}:|\vartheta|\le\frac{\pi}{6} \vee |\vartheta|\ge\frac{5}{6}\pi$. Thus we get
\begin{equation}
{\cal K}(t)=\frac{3a}{4\pi}\int_{\tilde{F}_{1/3}}\! d^2k\,\tilde{\rho}(\mathbf{k},t)\,k_x.
\end{equation}
Because the first \textit{K} point has coordinates $[\frac{4\pi}{3a},0]$ in the BZ, the valley index ${\cal K}\in[-1,1]$. ${\cal K}=1$ represents the \textit{K} valley, and ${\cal K}=-1$ represents the \textit{K'} valley. The valley index allows to track which valley state the electron occupies at a particular instant, namely follow the qubit state.

\subsection{NOT Gate}

The qubit has been defined on a pair of single electron conduction band states located in the \textit{K'} and \textit{K} valleys with spin up (1st and 3rd state).
We turn on the time simulations solving (Eq. \ref{ct}) self-consistently. 
The confinement potential is now modulated by applying oscillating voltages to two opposite gates $\mathrm{G}_1$ and $\mathrm{G}_3$ (see Fig.~\ref{fig:1}): $\mathrm{V}_{1,3}(t) = V_{dc}+ V_{ac}\left(\cos(\omega t)-1\right)$, $V_{ac} = 250$~mV, beside the constant negative bias $V_{dc} = -1500$~mV, which creates confinement. 
Voltages on the gates $\mathrm{G}_2$ and $\mathrm{G}_4$ remain constant: $V_{2,4}=V_{dc}$.
This modulation narrows the confining potential along one direction ($y$), which makes the electron state squeezed. The modulation angular frequency $\omega$ is tuned to the difference of energies $E_{KK'}=2\Delta\simeq3.5$~meV between two valley states, namely basis states of the qubit: $\omega=\omega_0=E_{KK'}/\hbar$.

\begin{figure}[t]
\includegraphics[width=3.8cm]{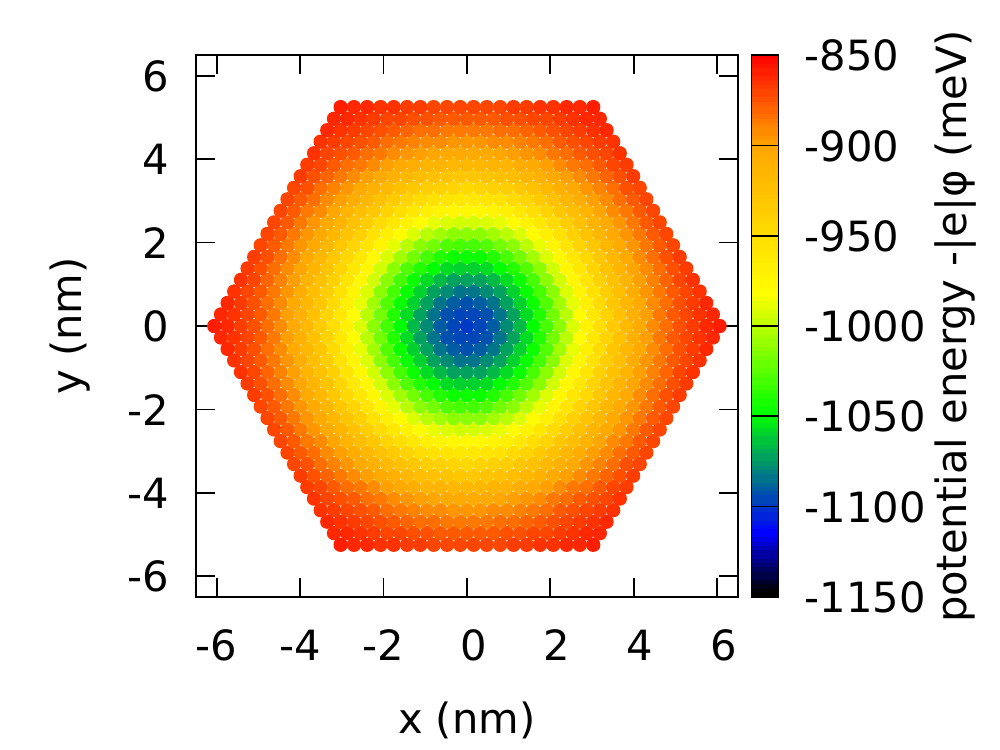}
\includegraphics[width=3.8cm]{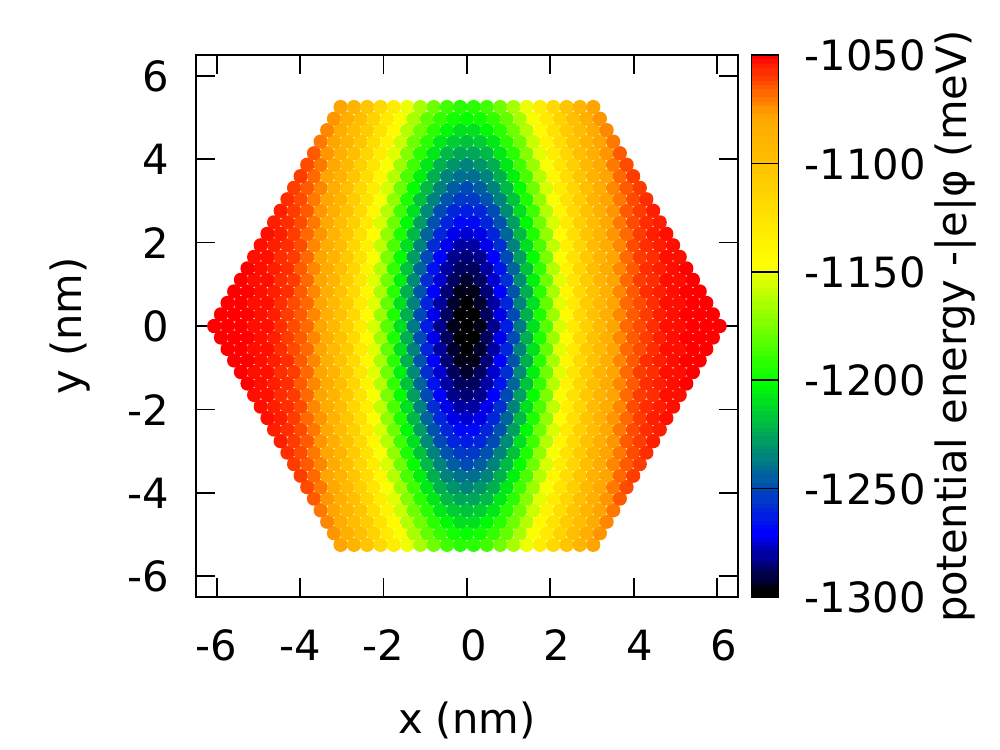}\\
\includegraphics[width=3.8cm]{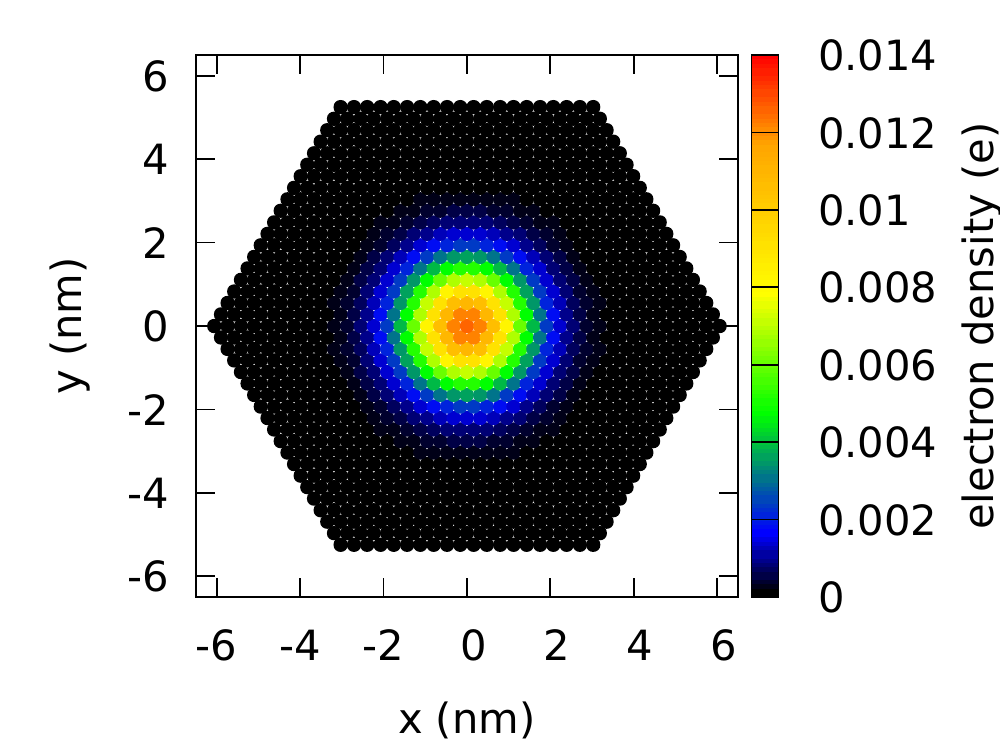}
\includegraphics[width=3.8cm]{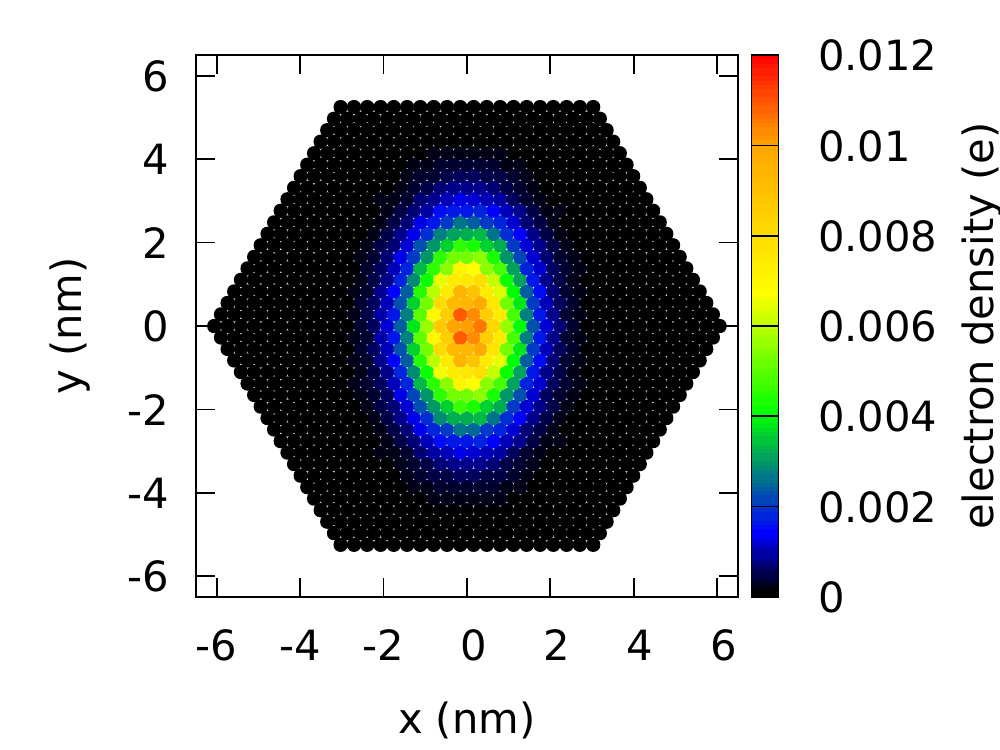}
\caption{\label{fig:7} The initial confinement potential (top left), and modulated (top right) at $t = T/2$, which causes squeezing of the electron density (bottom). The modulation is obtained by applying additional oscillating voltages to two opposite gates G$_1$ and G$_3$. }
\end{figure}

In the top left corner in Fig.~\ref{fig:7} we see a potential calculated for the initial instant $t=0$, when $V_{1,3}=-1500$~mV. Top left corner shows the potential at half a cycle for $t=T/2$, $T=2\pi/\omega$, namely at a time when the electron is squeezed the most. The initial electron density $\rho(\mathbf{r},t)$ is shown in the bottom left corner and at $t=T/2$ in the bottom right. It is clearly visible how narrowing of the potential squeezes the electron density.

\begin{figure}[h]
	\includegraphics[width=3.5cm]{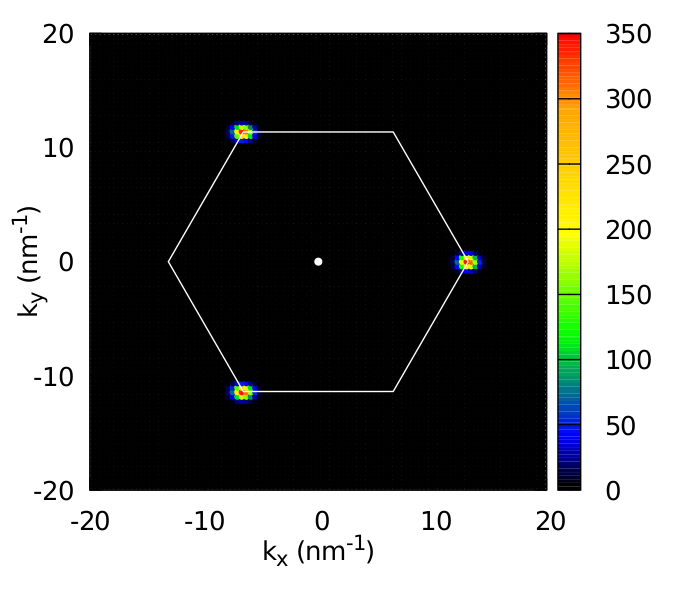}
	\includegraphics[width=3.5cm]{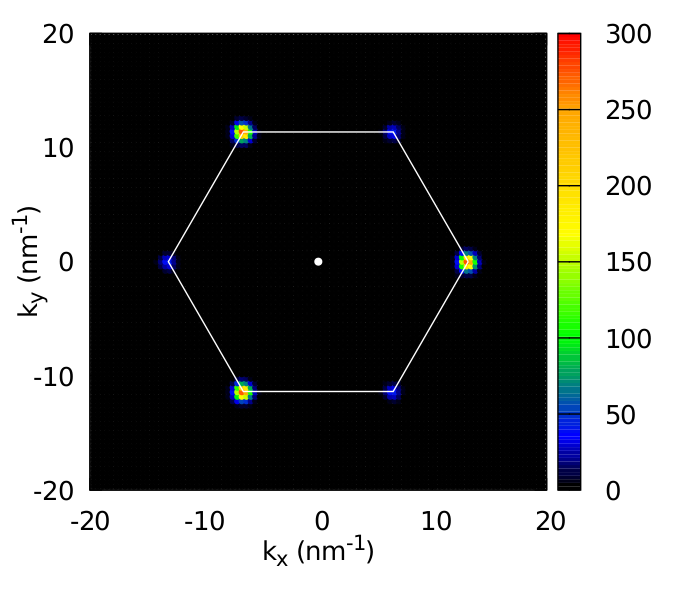}\\
	\includegraphics[width=3.5cm]{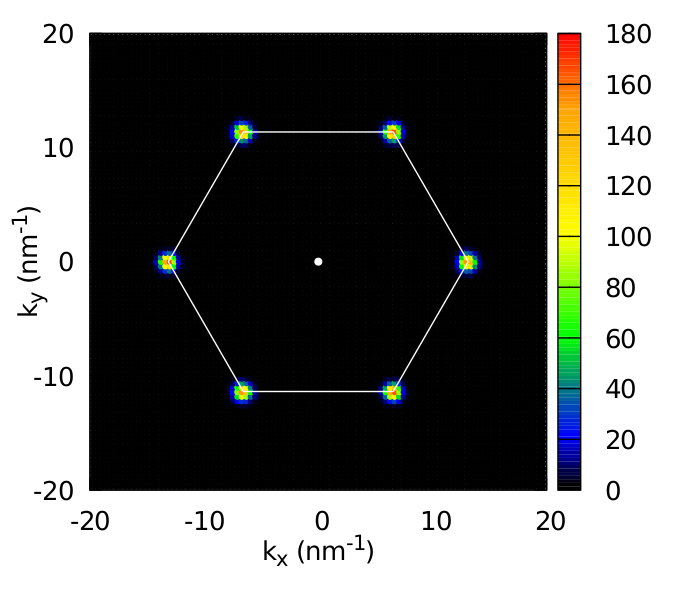}
	\includegraphics[width=3.5cm]{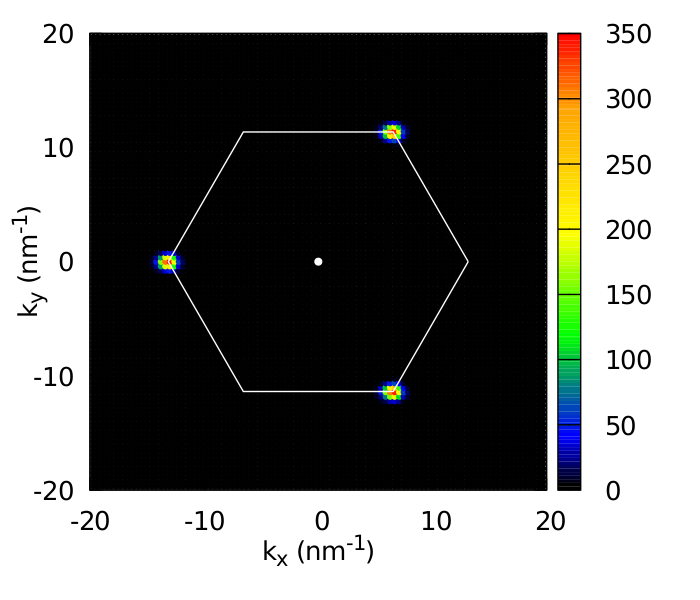}
	\caption{\label{fig:8} The electron density in the reciprocal space $\tilde{\rho}(k_x,k_y)$ calculated via Fourier transform of the electron state at $t=0$, 12, 26 and 52 (ps). During the evolution we observe gradual flow of the electron density from the valley \textit{K} to \textit{K'}.}
\end{figure}

Initially the electron occupies the \textit{K} valley state with valley index $\mathcal{K}(t=0)=1$. Oscillatory change of the electron density generates gradual transition of the electron from the \textit{K} valley to the \textit{K'} valley. In Fig.~\ref{fig:8} we see such an inter-valley transition. At time $t=26$~ps the electron is in an equally weighted linear combination of both valley states, while at twice that time ($t=52$) the transition is complete and the electron occupies the \textit{K'} valley state for which $\mathcal{K}(t =52\,\mathrm{ps})=-1$.

If we now calculate the Fourier transform $\Phi^{\sigma\alpha}(\mathbf{k},t)$ (Eq.~\ref{fout}) during the oscillations of the confinement potential, we see how this modulation induces gradual transition between valleys. Subsequent maps in Fig.~\ref{fig:8} show that the electron density inside the BZ gradually flows between the valleys.

\begin{figure}[b]
	\includegraphics[width=8.1cm]{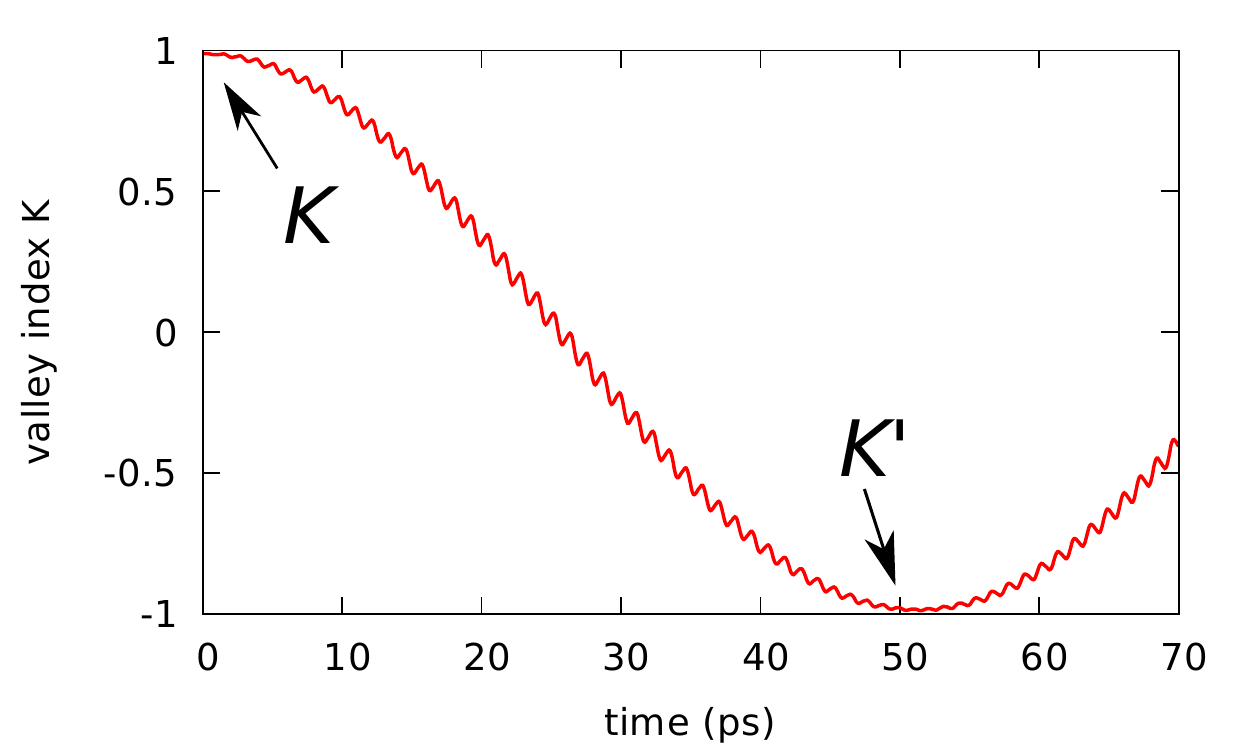}
	\caption{\label{fig:9} The transition between \textit{K} \& \textit{K'} valleys induced by the confinement potential modulation. The transition is equivalent to the valley-qubit NOT gate.}
\end{figure}

Time course of the valley index is shown in Fig.~\ref{fig:9}. Full transition between valleys is equivalent to performing the simplest NOT operation on a valley qubit: $\mathrm{NOT}\,| \mathcal{K}\! = \!1 \rangle=|\mathcal{K}\!=\!-1 \rangle$ at time $t=52$ (in general $\mathrm{NOT}\,|\,\mathcal{K}\,\rangle=|\!-\mathcal{K} \rangle$).
\linebreak Precisely speaking, we obtain an $\imath\mathrm{NOT}$ operation, as this represents a rotation on the Bloch sphere spanned by states $|\mathcal{K}\!=\!1 \rangle$ and $|\mathcal{K}\!=\!-1 \rangle$, which belongs to the special unitary group $\mathrm{SU}(2)$ of unitary matrices of size $2\times2$ and a determinant equal to 1 (NOT gate represented by $\sigma_x$ matrix has determinant $-1$). Moreover halfway through the $\imath\mathrm{NOT}$ gate operation, at $t=26$~ps, we obtain an operation $\mathrm{H} = e^{\imath\pi/4}\sqrt{\mathrm{NOT}}$ equivalent to the Hadamard operation, which is of high importance for quantum computing.
The $\mathrm{H}$ gate generates an equally weighted linear combination of opposite valley states $\mathrm{H}\,| \mathcal{K}\!=\!1 \rangle = \frac{1}{\sqrt{2}}\left(\imath|\mathcal{K}\!=\!1 \rangle + |\mathcal{K}\!=\!-1 \rangle\right)$. 

\begin{figure}[t]
	\includegraphics[width=8.3cm]{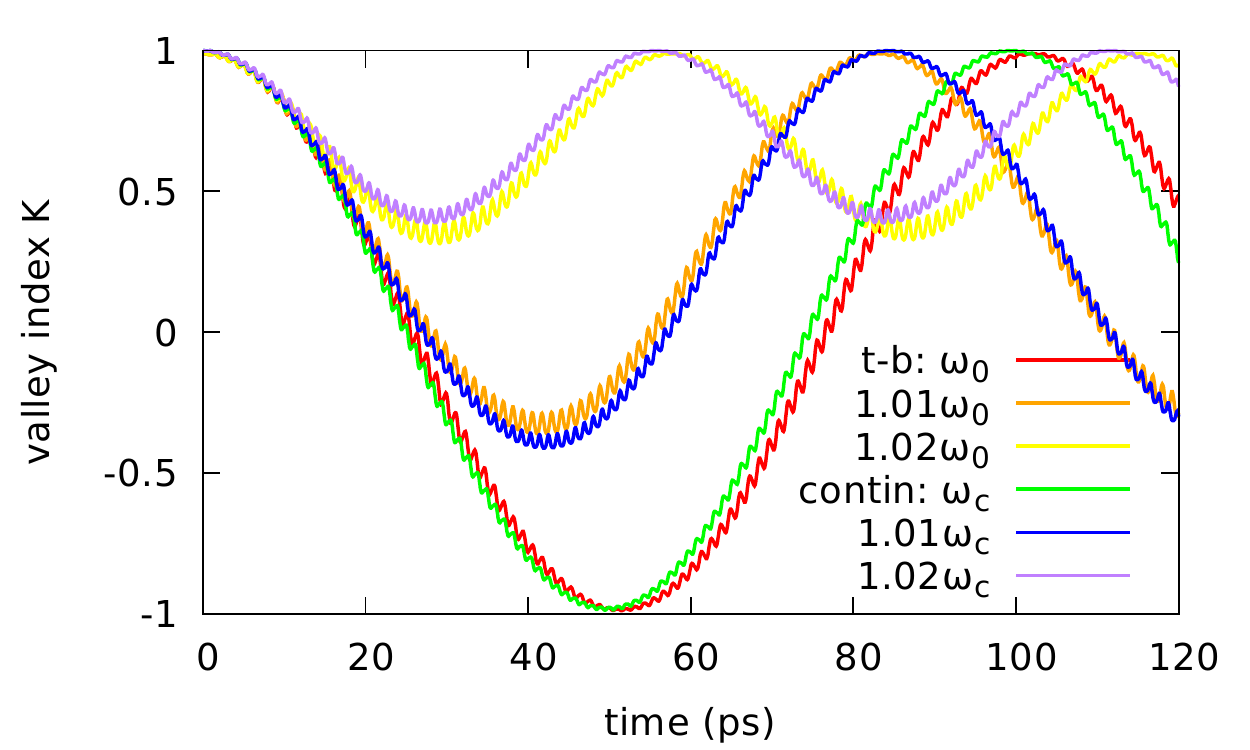}
	\caption{\label{fig:10} Transitions between valleys exhibit oscillations of Rabi type. For the resonant frequency $\omega=\omega_0$ (red curve) we obtain a full transition. Beyond the resonance frequency, as for the orange ($1.01\,\omega_0$) and yellow ($1.02\,\omega_0$) curves, we can observe incomplete transitions with amplitudes getting smaller as the distance from the resonance increases. For comparison, transition according to the continuum model for the same differences in frequencies are also shown as (the value of $\omega_c$ is slightly different than for TB): green curve (resonance $\omega_c$), blue curve ($1.01\,\omega_c$) and magenta curve ($1.02\,\omega_c$).}
\end{figure}

Transitions induced in the nanodevice are resonant. We emphasize that the pumping frequency $\omega$ must be tuned to the difference between the energies of states \textit{K} and \textit{K'}: $\omega=\omega_0$ ($=E_{KK'}/\hbar$). Any divergence from the resonant frequency results in incomplete Rabi oscillations. They get smaller as we move away from the resonant frequency $\omega_0$. Fig.~\ref{fig:10} shows resonant transitions (red) resulting in a full Rabi cycle $\omega=\omega_0$. Additional simulations were performed for frequencies diverging from the resonance. Incomplete cycles for $\omega=1.01\,\omega_0$ and $\omega=1.02\,\omega_0$ are marked as orange and yellow. 

\begin{figure}[t]
	\includegraphics[width=8.3cm]{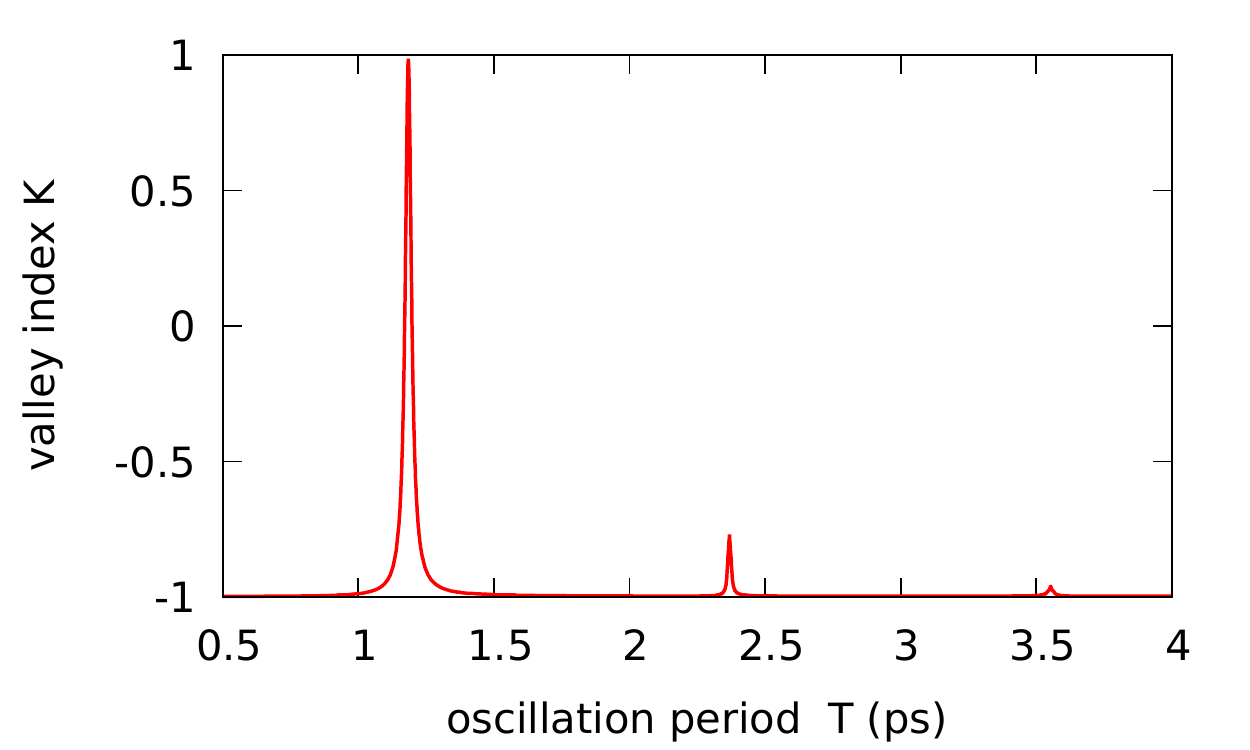}
	\caption{\label{fig:11} The transition between \textit{K} \& \textit{K'} valleys for different voltage oscillation periods $T=2\pi/\omega$. The resonant peak is clearly visible at $T_0 = 1.185$~ps. Additional fractional resonances are also present.}
\end{figure}

Now, let us assume that the electron is in an initial state $\mathcal{K}=-1$. If we now plot the maximal reached value of the valley index $\mathcal{K}(t)$ over a period of driving oscillations $T=2\pi/\omega$, we observe resonant peaks. This is shown in Fig.~\ref{fig:11}. For the resonant period $T_0=2\pi/\omega_0=1.185$~ps, we get the maximal value indicating a full inter-valley transition. If we diverge from the resonance, the maximal value of the valley index $\mathcal{K}(t)$ falls down rapidly and this is the area of incomplete Rabi cycles. Most interestingly we also observe fractional resonances \cite{fract} for lower frequencies at $\omega/2$ and (barely visible) $\omega/3$. They can be explained within the second-order perturbation theory \cite{osika}.

In [\onlinecite{17}] the authors show that the inter-valley coupling strongly depends on the size and symmetry of the confinement potential in the QD. By changing this potential we modulate coupling between valleys. This effect is next exploited to pump the transitions between valley states through oscillatory alteration of the width of the confinement potential along one direction ($x$). The inter-valley coupling can be achieved also in a different way. In nanotubes and monolayers the inter-valley transitions can be induced by the lattice disorder\cite{kou,palyi,kristen}, e.g., vacancies\cite{osika} or by adatoms adsorption\cite{chen}, that is, via modification of the confinement with peaks originating from the lattice defects. However, such sources of coupling are very difficult to control. We, on the other hand, modify the confinement in a perfectly controllable manner. One need only remember not to move the minimum of confinement over sulphur atoms, as this might lift the coupling\cite{17}, increasing the gate operation time.
Recently, tunable valley splitting induced by van der Waals interaction with substrate (hBN) was presented in graphene quantum dots\cite{morgen2018}. Interestingly, it can be controlled electrostatically by displacements of the dot potential minimum.

In the course of transitions in Figs. 9 and 10 we can also notice a minor oscillation structure of frequency $\omega$ connected to a single cycle of pumping and oscillations of voltages applied to gates $\mathrm{G}_1$ and $\mathrm{G}_3$. These small oscillations of the valley index might affect fidelity of operations performed on the qubit. Fortunately, as we get closer to the basis states (poles on the Bloch sphere) their amplitude decreases. Moreover, we can reduce them arbitrarily by decreasing the amplitude $V_{ac}$ of voltage oscillations. However, to perform a full transition between valleys we need more cycles, which effectively extends the operation time.
For example to obtain an error of the order of 1\% during the NOT gate operation from Fig.~\ref{fig:9} we need $\sim50$ pumping cycles.

The duration of operations on qubits should be much shorter than the coherence time. 
Recently a long lifetime of an exciton valley state has been observed in a TMDC heterostructure, reaching 40~ns\cite{long}. These times are three orders of magnitude longer than the duration of our NOT operation.
The necessary frequencies of modulation of voltages in the range of several hundreds GHz might be problematic to obtain experimentally. One solution to lower these frequencies would be finding another material of a slightly lower spin-orbit splitting $2\Delta$. 
Another solution could be redefining the qubit using states $|\mathcal{K},s \rangle = |\!-\!1,\uparrow\,\rangle$ and $|\,1,\,\downarrow\,\rangle$, or, equivalently, the second degenerate pair: $|\,1,\,\uparrow\,\rangle$ and $|\!-\!1,\downarrow\,\rangle$.\cite{4} 
The energy splitting of such two-level system could be tuned using an external magnetic field. However, to obtain a coupling between the spin and valley degrees of freedom a spin-dependent confinement potential is necessary\cite{var3}. In our case the potential modulation could be carried out in a spatially variable magnetic field.

We should note however, that if we need an error not greater than 1\% (and for this we need several dozen pumping cycles per operation) and a number of operations during the qubit lifetime of about $10^3$, the pumping frequency must be of hundreds of GHz order.

\subsection{Qubit Readout}
Each full qubit implementation, aside from a possibility of performing operations, has to allow for qubit initialization and readout. To read out, as well as to initialize a valley qubit we can utilize the spin and valley Pauli blockade, so far observed in carbon nanotubes\cite{kou,laird,osik3}. The blockade imposes selection rules, which block transport of electrons of the same spin and valley as the electron in an adjacent quantum dot.

Let us assume that adjacent to the dot, in which our qubit is defined, we put another dot on a right with an electron initialized in the ground state in the \textit{K'} valley and with spin up, i.e. $|\mathcal{K},s \rangle = |\!-\!1,\uparrow\,\rangle$. Assuming that valley and spin are conserved during tunneling, the electron representing our qubit cannot tunnel to the right dot if the electron there occupies the same valley and spin state. We thus have a valley blockade $|\!-\!1,\uparrow\,\rangle(1,1)|\!-\!1,\uparrow\,\rangle$.\cite{osik4,burk}
However if, in the course of operation, we perform a valley transition, then the blockade is lifted and the electron can freely enter the right dot: $|\,1,\,\uparrow\,\rangle(1,1)|\!-\!1,\uparrow\,\rangle$ $\rightarrow$ 
$(0,2)|\,1,\,\uparrow\,\rangle|\!-\!1,\uparrow\,\rangle$. 
In this way, by extending the system with an adjacent dot trapping a reference electron with given valley state, we can perform a valley qubit readout. Nonetheless, this method still requires experimental confirmation in TMDC materials.

In real situations during electron tunelling between the dots an inter-valley mixing might take place, which results in a readout error. There are two sources of such mixing. The first is the tunelling itself through the barrier between dots, which magnitude can be assumed as the same order of magnitude as for the inter-valley coupling during the potential squeezing in our method, namely $\sim40~\mathrm{\mu eV}$. The second is mixing induced by point defects and dislocations. It indeed vanishes at the clean limit, however in practice, the available samples so far, are quite disordered. Let us assume that for a typical defect concentration of $0.1$~nm$^{-2}$,\cite{hong} the disorder-induced valley mixing energy scale is $50~\mathrm{\mu eV}$.\cite{paly} This gives an about 0.5\% admixture of the opposite valley-state\cite{pearce} during tunneling, and the same readout error. Eventually, together with mixing due to passing through a barrier during tunneling between dots, this gives the readout error estimate of about 1\%.

\section{Continuum Model}
There are numerous models within the $\mathbf{k}\cdot\mathbf{p}$ approximation in TMDCs, which successfully describe the electron dynamics near the conduction band minimum\cite{4,17,18,19,20}. In such continuum models we neglect the details of the microscopic lattice-periodic crystal potential. We now introduce a continuum model taking into account the ground state level in the conduction band at the minimum \textit{K}(\textit{K'}), with spin and valley degeneracy. This level is further split by $2\Delta$ by the intrinsic spin-orbit interaction. Within this model we include four subbands with states
$\{|\mathcal{K},s \rangle=|s\rangle\otimes|\mathcal{K}\rangle\}=\{|\,1,\uparrow\, \rangle,|\!-\!1,\uparrow\,\rangle,|\,1,\downarrow\,\rangle,|\!-\!1,\downarrow\,\rangle \}$, forming a basis. The model is introduced according to [\onlinecite{4}] and [\onlinecite{17}].

The model Hamiltonian includes the effective masses $m^{\mathcal{K},s}$ for a set of four states $|\mathcal{K},s \rangle$, with $m^{-\mathcal{K},s}=m^{\mathcal{K},-s}$. The assumed band masses are $m^{1\uparrow} = m^{-1\downarrow} = 0.49$~$m_e$, $m^{-1\uparrow} = m^{1\downarrow} = 0.44$~$m_e$.\cite{4} The resulting $4\times4$ Hamiltonian matrix assumes the following form:
\begin{equation}\label{cont}
 \begin{pmatrix}
  \frac{\hbar^2 k^2}{2m^{1\uparrow}}+\Delta & \Lambda & \lambda_R^\ast k_{-} & 0 \\
  \Lambda &  \frac{\hbar^2 k^2}{2m^{-1\uparrow}}-\Delta & 0 & \lambda_R^\ast k_{-} \\
  \lambda_R k_{+} & 0 & \frac{\hbar^2 k^2}{2m^{1\downarrow}}-\Delta & \Lambda \\
  0 & \lambda_R k_{+} & \Lambda & \frac{\hbar^2 k^2}{2m^{-1\downarrow}}+\Delta
\end{pmatrix},
\end{equation}
with square of the momentum operator $\hbar^2k^2=\hbar^2(k_x^2+k_y^2)$, $k_{\pm}=k_x\pm\imath k_y$, and $k_i\equiv-\imath\partial_i$. The full wave function (envelope) in this representation takes the form $\Psi(\mathbf{r}) = [\psi_{1\uparrow}(\mathbf{r}),\psi_{-1\uparrow}(\mathbf{r}),\psi_{1\downarrow}(\mathbf{r}),\psi_{-1\downarrow}(\mathbf{r})]^\intercal$.
This model allows to include the Rashba type spin-orbit interaction, induced by an external perpendicular electric field $E_z$ and with (complex in general) spin orbit coupling $|\lambda_R|=\gamma_R|e|E_z$, where $\gamma_R=3.3\cdot10^{-4}$~nm$^2$ and the field is expressed in $(E_z)=$~V~nm$^{-1}$.\cite{4} We assume a real value for the $\lambda_R=\gamma_R|e|E_z$ and the electric field of the order of $E_z \sim 50$~mV~nm$^{-1}$. However the form of the Hamiltonian (Eq. \ref{cont}) implies that the Rashba coupling has no influence on inter-valley transitions (it only couples the spin degrees of freedom), what has been confirmed by our numerical simulations.

Having the effective model introduced, we now want to calculate the evolution of the valley qubit, as we did for the tight-binding model, and compare the results. This way we can confront the TB simulations with another model and, and the same time, verify the continuum model. Beside the $2\Delta$ spin splitting in CB, to define the model we need to specify the strength of the inter-valley coupling $\Lambda$. The $2\Delta$ has been obtained from the TB model, by calculating splitting for the initial ($t=0$) form of the confinement potential and in the state of maximal squeeze $t = T/2$. We concluded that $2\Delta$ changes from $3.72$ to $3.66$~meV, during the potential pulsing. Thus in the continuum model we assume temporal variability $\Delta(t)=\Delta_1(\cos(\omega t)+1)/2+\Delta_0$ and parameters $\Delta_0=1.83$~meV, $\Delta_1=0.03$~meV. While the doubled inter-valley coupling $2\Lambda$ can be calculated from the difference between the ground state and the 1st excited state in CB under an assumption of no SOI interaction\cite{17,18}. We again calculate them from the TB approximation, and conclude that $2\Lambda$ changes from 0.088 to 0.008. For the purpose of the continuum model we assume $\Lambda(t)=\Lambda_1(\cos(\omega t)+1)/2+\Lambda_0$, $\Lambda_0=0.004$~meV, $\Lambda_1=0.04~$meV.

For the Hamiltonian (Eq.~\ref{cont}) we solve the time-dependent Schr\"odinger equation with varying $\Delta(t)$ and $\Lambda(t)$. The results are shown in Fig.~\ref{fig:10}. The green curve presents results for a resonant driving $\omega=\omega_c$, and for deviations from the resonance: $\omega=1.01\,\omega_c$ (blue) and $\omega=1.02\,\omega_c$ (magenta). The resonant frequency $\omega_c$ is slightly different, corresponding to $T_c=2\pi/\omega_c=1.099$~ps. The evolution within the continuum model is quite consistent with the tight-binding calculations. This speaks in favour of the continuum model for describing electron dynamics at the edge of the conduction band. It requires a significantly lower number of parameters than TB, yet still the values and ranges of parameters (ie. $\Delta_1$, $\Delta_2$, $\Lambda_1$, $\Lambda_2$) have to be specified for a particular form of the confinement potential. These values can be obtained from the TB model.

Despite somewhat different resonant frequencies between the TB and continuum models, the frequency and the amplitude of Rabi oscillations are very close to each other in both cases. This is because near the resonance, the Rabi frequency $\Omega_{0,c}=\sqrt{(\omega\!-\!\omega_{0,c})^2+(\Lambda_1/\hbar)^2}$ weakly depends on the driving oscillation frequency $\omega$, while it depends linearly on the amplitude $\Lambda_1$ of the inter-valley coupling oscillations. 

A natural requirement for a boundary is a condition that the normal probability current $\mathbf{j}(\mathbf{r})$ at the edge vanishes\cite{20}: $j_\bot=0$. 
In the nanodevice we use strongly bound states in the center of the flake, hence the boundary effects are unimportant for us and we can enlarge the flake and move its edges arbitrarily far away from the confinement center. That is why in the continuum model we assumed a flake of a much larger size than the confinement region and assumed a simplified square shape of the flake. At the boundary of the computational square we put $\nabla_{\!\bot} \Psi=0$, which automatically meets the boundary condition $j_\bot=0$.

\section{Conclusions}

We investigated a prospect of a valley-qubit realization presenting results of computer simulations in which we perform operations on a valley-qubit defined in a nanodevice based on a MoS$_2$ monolayer. 
Such monolayers have a number of interesting properties. Unlike graphene, they have a band gap and a relatively high value of the spin-orbit coupling. Moreover, they possess an ability of using the valley pseudospin, in addition to spin, as a quantum bit. Thus, they are important materials for spintronics and recently introduced valleytronics. 
We described the monolayer within the tight-binding approximation using a basis of three molybdenum orbitals (three-band tight-binding). This model includes an external confinement potential, which traps an electron, created by voltages applied to local gates. The exact form of the dot potential is obtained by solving the Poisson equation for the presented gates layout, the MoS$_2$ flake and separating dielectric layers.

In the quantum dot we trap a single electron. We modulate the confinement potential by applying oscillatory voltages to two opposite gates, effectively squeezing the electron wave function. The oscillating voltages induce gradual transitions of the electron state from the \textit{K} to \textit{K'} valley (the spin remains constant). The transitions between valleys are identical to the operation of the elementary quantum NOT gate on the valley qubit. These transitions have the Rabi oscillations form. 

Simulation of this kind are carried out by solving the time-dependent Schr\"odinger's equation in a basis of tight-binding Hamiltonian eigenstates for a time-dependent confinement potential, calculated along with the Poisson's equation at every time step. The tight-binding calculations are further confirmed by simulations within the 4-band continuum model, with good consistency. At the same time we validate the continuum models and show their usefulness in describing time-evolutions of conduction band electron states in TMDC monolayers.

As a result of the performed simulations, we showed feasibility of operations on a valley-qubit implemented in modern monolayer materials. We have developed, previously used\cite{5,5a}, precise and realistic simulations of the time evolution of semiconductor nanodevices. Now introduced for new and attractive materials with interesting properties.

\begin{acknowledgments}
JP would like to thank Pawe\l{} Potasz for invaluable discussions, and Grzegorz Skowron for help with the article editing.
This work has been supported by National Science Centre (NSC), Poland, under Grant No. 2016/20/S/ST3/00141.
SB acknowledge support from NSC, Grant No. UMO-2014/13/B/ST3/04526.
This research was supported in part by PL-Grid Infrastructure.
\end{acknowledgments}



\appendix*
\section{Dispersion relation for a finite flake}

\begin{figure}[t]
	\includegraphics[width=8.3cm]{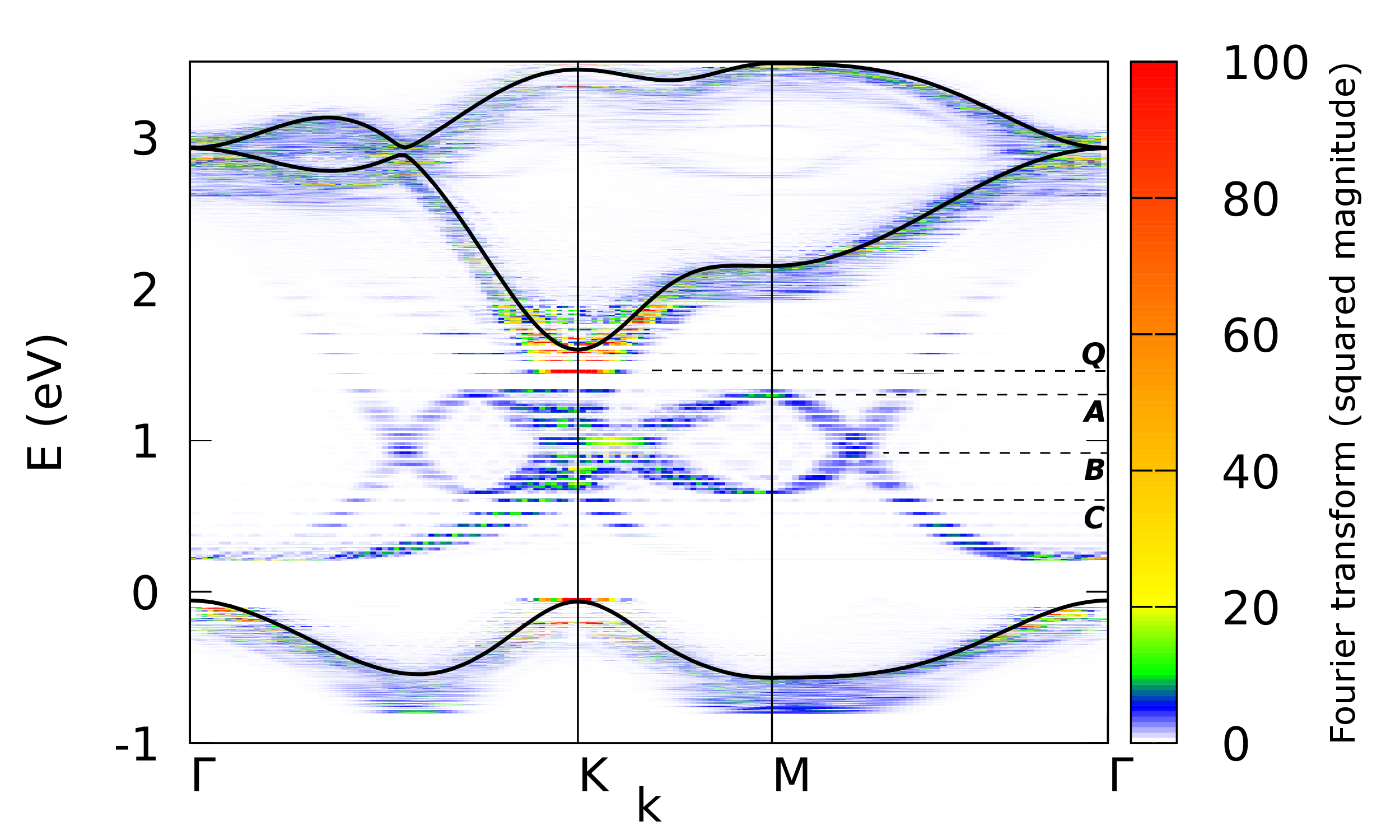}
	\caption{\label{fig:12} Dispersion relation $E(k)$ for a finite flake MoS$_2$ obtained through plotting the density $\tilde{\rho}_m(k)$ along a symmetry axis in the BZ for subsequent energies $E_m$. Obtained this way bands recreate successfully the dispersion relations for an infinite flake (marked as black lines) within the 3-band model. Moreover, sub-bands of trivial edge states are visible, which do not close the band gap.}
\end{figure}

Let us get back to Fig.~\ref{fig:6} showing electron densities in the reciprocal space $\tilde{\rho}_m(\mathbf{k})=|\Phi_m(\mathbf{k)}|^2$ obtained from Fourier transforms of the eigenstates $\boldsymbol{\psi}_m(\mathbf{r})$. Now, we plot in color subsequent densities $\tilde{\rho}_m(k)$ along lines connecting points of high symmetry $\Gamma$-$K$-$M$-$\Gamma$ in the BZ (obtaining, this way, 1D densities along these directions) at an energy level $E_m$ corresponding to subsequent states. Thus forming 2D color maps in the wave vector space $(k,E(k))$. This way we get colored dispersion relations $E(k)$ shown in Fig.~\ref{fig:12}. Additionally three black curves show the relations $E^{\mathrm{inf}}(k)$ obtained using our 3-band model (Eq.~\ref{ham1}) for a case of an infinite monolayer. We see that the valence and conduction bands are formed with a direct band gap at the point $K$.
The obtained sub-bands $E(k)$ faithfully recreate the dispersion relations of an infinite flake $E^{\mathrm{inf}}(k)$. The visible widening (blurring of the dispersion relations) clearly results from a finite size of the modeled flake. This way we verified correctness of the obtained eigenstates in the flake. A characteristic feature is that the bound states in the dot (especially those from the bottom of the continuum band near the point $K$) are slightly shifted towards lower energies, which is typical for bound systems. The dashed line Q in Fig.~\ref{fig:12} marks the states from the bottom of the conduction band which are used to define our qubit. Notice, that subsequent (quantized!) states lie here increasingly closer to each other on the energy scale, which is typical for bound states in a finite well.

Nevertheless, the most interesting result of our method of computing the $E(k)$ relation for a finite flake is obtaining branches of edge states. In Fig.~\ref{fig:12} we see additional sub-bands in the energy gap, created by the edge states. Lines A, B and C are used to mark densities for edge states from the second row of Fig.~\ref{fig:6} with their corresponding eigenenergies. A notable feature is the fact that these states do not close the energy gap (it is still visible in the vicinity of the valence band), which means that the edge states are topologically trivial\cite{topo1,topo2}. They start in the conduction band but do not reach the valence band and are still gapped.


\bibliography{thebibliography}

\begin{thebibliography}{82}
\bibitem [{not()}]{note1}%
  \BibitemOpen
  \href@noop {} {}\bibinfo {note} {However, due to symmetry of the voltage
  changes in time (oscillations) we can take only half of the cycle
  $T$.}\BibitemShut {Stop}%
\end{thebibliography}



\end{document}